\documentclass[a4
paper, 10pt]{elsarticle}

\usepackage[english]{babel}
\usepackage{babel}
\usepackage{fontenc}
\usepackage{graphicx}
\usepackage{amsmath,amsthm,amssymb}
\usepackage{natbib}
\usepackage[colorlinks=true,citecolor=blue]{hyperref}
\usepackage{color}
\usepackage{graphicx}
\usepackage{subfigure}
\usepackage{color}
\usepackage{newlfont}
\usepackage{multirow}
\usepackage{longtable} 
\usepackage{ifthen}
\usepackage{alltt}
\usepackage{enumerate}
 \usepackage[text={6.25in,8.5in},centering]{geometry}
\usepackage{tikz}
\newcommand{\ba}{\begin{array} }
\newcommand{\ea}{\end{array} }
\newcommand{\bae}{\begin{eqnarray}}
\newcommand{\eae}{\end{eqnarray}}
\newcommand{\bea}{\begin{eqnarray*}}
\newcommand{\eea}{\end{eqnarray*}}
\newcommand{\be}{\begin{equation}}
\newcommand{\ee}{\end{equation}}

\usepackage{amssymb}
\usepackage{amsthm}
\usepackage{float}
\usepackage{amsmath}

\usepackage{tikz}
\usetikzlibrary{shapes,arrows}
\numberwithin{equation}{section}
\newtheorem{theorem}{Theorem}[section]
\newtheorem{lemma}{Lemma}[section]

\newtheorem{remark}{Remark}[section]

\input epsf.sty
\usepackage{lineno}

\journal{}

\begin{document}
\title{Effect of active case finding on dengue control: Implications from a mathematical model}
\author[ISI]{Indrajit Ghosh \footnote{Corresponding author. Email: indra7math@gmail.com}}
\author[KU]{Pankaj Kumar Tiwari}
\author[ISI]{Joydev Chattopadhyay}
\address[ISI]{Agricultural and Ecological Research Unit, Indian Statistical Institute, Kolkata - 700 108, West Bengal, India}
\address[KU]{Department of Mathematics, University of Kalyani, Kalyani - 741 235, West Bengal, India}

\begin{abstract}
Dengue control in India is a challenging task due to complex
healthcare settings. In yesteryears, an amplification of dengue
infections in India posed the need for introspection of existing
dengue control policies. Prior understanding of the impacts of
control interventions is necessary for their future implementation.
In this paper, we propose and analyze a compartmental model of
dengue to assess the impact of active case finding (ACF) on dengue disease
transmission. Currently, primary prevention of dengue is possible only with vector control and personal protection from the
bites of infected mosquitoes. Although a few experimental studies are performed to assess ACF in dengue disease, but this is the first attempt to represent and study the dynamics of disease using ACF as a control strategy.
Local and global dynamics of the system are studied. We use
sensitivity analysis to see the effects of controllable parameters
of the model on the basic reproduction number and total number of
infective population. We find that decrease in the biting rate of
mosquitoes, and increase in the rate of hospitalization and/or
notification, death rate of mosquitoes and ACF for
asymptomatic and symptomatic individuals play crucial role for the
reduction of disease prevalence. We calibrate our model to the
yearly dengue cases in eight dengue endemic states of India. The
results of our study show that ACF of symptomatic
individuals will have significant effect on dengue case reduction
but ACF of asymptomatic individuals cannot be
ignored. Our findings indicate that the healthcare organizations
must focus on ACF of symptomatic as well as
asymptomatic individuals along with personal protection and
mosquitoes control to achieve rapid reduction of dengue cases in
India.

\end{abstract}

\maketitle

\section{Introduction}\label{sec1}
Dengue fever is a vector borne disease, transmitted by the bite of Aedes aegypti or
Aedes albopictus female mosquitoes, distributed mainly in tropical and subtropical
areas and is caused by four closely related dengue serotypes (DENV
1-4) \cite{Guzman2002,Bancroft1906}. Almost all age groups can be
affected by dengue. Its symptoms become apparent after 3–-14 days of
the bite of infected mosquito \cite{WHO2013a}. After recovery from
one serotype of dengue, one can become immune to that particular
serotype, but is prone to get infection with the remaining three
serotypes \cite{Gubler1998}. In recent years, the rate of dengue
cases is accelerating and figures from World Health Organization
confirm 284--528 million cases per year around the globe
\cite{WHO2013}. Unfortunately, no effective vaccine is available
against dengue fever \cite{Whitehorn2010}, but there are numerous
candidate vaccines (such as, inactivated whole virus vaccines, live
attenuated mono and tetravalent formulation, and recombinant subunit
vaccines), undergoing clinical trials in various phases
\cite{Blaney2005,CDC2007,CVD2007,Dung1999,Mutheneni2017}. Thus, in
order to avoid dengue infection, people have to prevent themselves
from the mosquitoes. Use of insecticide, removal of mosquito
breeding sites generated by humans in households (e.g., old toys,
water containers, and tires) and making people aware about mosquito
net and other mosquito repeller are some useful ways for controlling
the disease \cite{WHO2013,Gratz1991}. However, these methods are not
sufficiently effective due to frequent outbreaks of the disease in
some areas. In such scenarios, mathematical models of vector-borne
diseases include ideas of how to curb the disease.

After the seminal work of Fischer and Halstead \cite{Fischer1970},
research on dengue transmission has gained commendable attention. In
most of the studies, effect of epidemiological interactions between
multiple dengue serotypes has been studied. Feng et al.
\cite{Feng1997} examined the principle of competitive exclusion in a
dengue model consisting of two serotypes. Using spatial epidemic
data, basic reproduction number for a dengue model was estimated by
Chowell et al. \cite{Chowell2007}. Tewa et al. \cite{Tewa2007}
studied a dengue model by considering one type of virus. Using
Lyapunov functions, they have shown global asymptotic stability of
disease-free and endemic equilibria. Garba et al. \cite{Garba2008}
developed and rigorously analyzed a compartmental model for spread
and control of dengue disease. They considered transmission by
exposed humans and mosquitoes as well. They showed that the model
exhibits the phenomenon of backward bifurcation when standard
incidence is considered. Further, they have shown that taking into
account mass-action incidence function, no backward-bifurcation can
be observed. Derouich et al. \cite{Derouich2003} have shown that
employing environment management or chemical methods for prevention
of dengue is not sufficient; it can only delay the outbreak of the
disease. Influence of spatial heterogeneity on the disease emergence
has been explored by Favier et al. \cite{Favier2005}. Some
stochastic models for dengue infection are also available in
literature \cite{Focks1995,Medeiros2011,Nuraini2012}. Degallier et
al. \cite{Degallier2009} and Kongnuy et al. \cite{Kongnuy2011} have
investigated the dynamics of dengue by using statistical methods.
Using numerical techniques, Perez et al. \cite{Perez2009}
investigated the dynamics of dengue disease. Some mathematical
models have been studied for dynamics and control of dengue
infection
\cite{Chowell2007,Garba2008,Dietz1975,Esteva2005,Newton1992}. In
some models, optimal control theory is used to design the paths to
limit the spread of dengue \cite{Thome2010,Antonio2001}.

To the best of our knowledge, the aforementioned studies have
focused on mosquito control and personal protection as preventive
measures, but didn't attempt to assess the impact of active case
finding as a control measure for the dengue epidemic. Active case
finding (ACF) requires a special effort by the healthcare
organizations to increase the detection of dengue in a given
population. This strategy identifies and brings into treatment
people with dengue who have not sought diagnostic services
themselves. ACF can reduce the number of subsequent dengue
infections and prevent secondary cases by detecting and treating
patients on the early stage of infection. ACF is conducted by
trained healthcare agents who make face-to-face contact with
patients and immediately prescribes onsite evaluation
\cite{Golub2005,Mandal2015,Corbett2010,Ruche2010}. Moreover, ACF can
be used to fill the data gaps caused by under reporting of dengue
cases. A community-based active dengue fever surveillance among the
individuals of age group 0--19 years has been conducted by Vong et
al. \cite{Vong2010}. The surveillance is done in rural and urban
areas of Cambodia, Kampong Cham, during 2006-2008. The main purpose
of surveillance was to estimate the true burden of dengue in the
area. As a result of this surveillance, a higher disease incidence
was found than that reported to the national surveillance system.
Moreover, the incidence of disease was found to be high in both
rural and urban areas, especially in preschool children. In a more
recent study \cite{Vikram2016}, authors attempted to quantify the
proportion of asymptomatic dengue infection in some regions of
Delhi. They tested a total of 2,125 persons, with or without
symptoms of dengue and found relatively high prevalence of
asymptomatic cases to that of symptomatic cases. Further,
symptomatic dengue patients were referred to nearby hospitals for
medications while cases with asymptomatic infections were provided
necessary knowledge about subsequent secondary infections. This
study is somehow similar in nature to that of ACF. Thus, ACF can be
helpful to reduce dengue burden in two ways: it will fill the data
gaps caused by under-reporting and simultaneously detection of
dengue patients in early stage will prevent secondary infections.
This twofold benefits of ACF motivated us to investigate its impact
on dengue control. As far our knowledge goes, this is the first
attempt to represent the dengue dynamics mathematically using ACF as
a control strategy.

For case study, we consider dengue prevalence during 2007--2017 in
eight dengue endemic states of India, namely: Kerala, Delhi,
Gujarat, West Bengal, Andhra Pradesh, Rajasthan, Maharashtra and
Karnataka. In India, the dengue haemorrhage fever/dengue shock
syndrome (DHF/DSS) occurred in various parts of the country since
1988 and major outbreak was occurred around Delhi and Lucknow in the
year 1996 \cite{Dar1999}. India experienced significantly high
levels of dengue cases during last two decades
\cite{Chakravarti2012}. In 2018, a provisional total of 14,233
dengue cases (with 30 deaths) has been reported in India till 22nd
July \cite{NVBDCP2018}. In order to stop the rising number of dengue
cases, the government of India has launched a major campaign to
enhance awareness among people about methods of prevention. Despite
the use of various methods for the control of mosquitoes and
personal protections by people themselves, dengue is still not under
control in most of the states.

Remainder of the paper is presented in the following way: Section \ref{model}
contains model formulation and underlying assumptions. Dynamics
of disease-free equilibrium is studied in Section
\ref{disease_free}. We analyze the system for feasibility and
stability of endemic equilibrium in Section \ref{endemic}.
Sensitivity analysis is performed in Section \ref{sensitivity1}. We
calibrate our model for yearly dengue case data of eight different
states of India in Section \ref{estimation}. In Section
\ref{control}, we study the impacts of ACF on dengue burden in these
states. The paper ends with discussion and conclusion in Section \ref{conclusion}.

\section{Model formulation}\label{model}

At any time $t>0$, the total mosquito population ($N_v$) is
sub-divided into three classes: susceptible mosquitoes ($S_v$),
mosquitoes exposed to the dengue virus ($E_v$) and infected
mosquitoes ($I_v$). We divide the total human population ($N_h$)
into seven sub-classes: high risk susceptible individuals
($S_{h1}$), low risk susceptible individuals ($S_{h2}$), individuals
exposed to dengue virus ($E_h$), asymptomatic individuals ($A_h$),
individuals with dengue symptoms ($I_h$), hospitalized and/or
notified individuals ($P_h$) and recovered individuals ($R_h$).
Symptomatic dengue infection is referred to fever with atleast two
symptoms of dengue (headache, retro-ocular pain, arthralgia, myalgia
and rash) while asymptomatic dengue infection is defined as no
clinical symptoms of dengue as in the case of symptomatic
infection \cite{WHO2013}. The individuals in the $P_h$ class are
assumed be those who are admitted to the hospital and the people who
are notified as confirmed dengue patients.

Susceptible mosquitoes are assumed to be recruited at a constant
rate $\pi_v$. They move to the exposed class by acquiring dengue
through contacts with infected humans (asymptomatic and
symptomatic). We consider standard incidence for the interactions
between susceptible mosquitoes and infected humans. Exposed
mosquitoes are assumed to move to the infected class at a rate
$\gamma_v$. Mosquitoes in each class are assumed to die from natural
causes at a rate $\mu_v$. Recovered class is not considered for the
mosquito population. The reason behind this is once infected from
dengue virus, the female mosquitos remain infected throughout their
life span \cite{Chamberlain1961}.

The individuals are recruited in the region at a constant rate
$\pi_h$ (by birth or immigration) and assumed to join the
susceptible class. A fraction $r$ of total newly recruited
populations join the high risk susceptible population ($S_{h1}$) and
the remaining join the low risk susceptible population ($S_{h2}$).
Individuals in classes $S_{h1}$ and $S_{h2}$ are assumed to join the
exposed class on effective contact with infected mosquitoes. The
interactions between susceptible humans and infected mosquitoes are
assumed to be of standard incidence type. The individuals in class
$S_{h2}$ are assumed to contract the disease at a lower rate than
the individuals in class $S_{h1}$. The exposed humans are assumed to
move in the infected class at a rate $\gamma_h$, a fraction $\rho$
of which join the asymptomatic class $A_h$, while the remaining ones
join the symptomatic class $I_h$. The individuals in the symptomatic
class are hospitalized and/or notified at a constant rate $\eta$.
The individuals in classes $A_h$, $I_h$ and $P_h$ are assumed to
recover from the disease at the rates $q_1$, $q_2$ and $q_3$,
respectively. The natural death of individuals in each class is
assumed to be at a constant rate $\mu_h$. Furthermore, the
asymptomatic and symptomatic individuals are notified through ACF at
constant rates $p_1$ and $p_2$, respectively. The recovered
individuals do not acquire the infection again as they get lifelong
immunity. We assumed that the hospitalized and/or notified
individuals are not going to infect others because they will be kept
in mosquito-free environments.
\begin{figure}[!t]
   \centering
   \includegraphics[height=10cm,width=15cm]{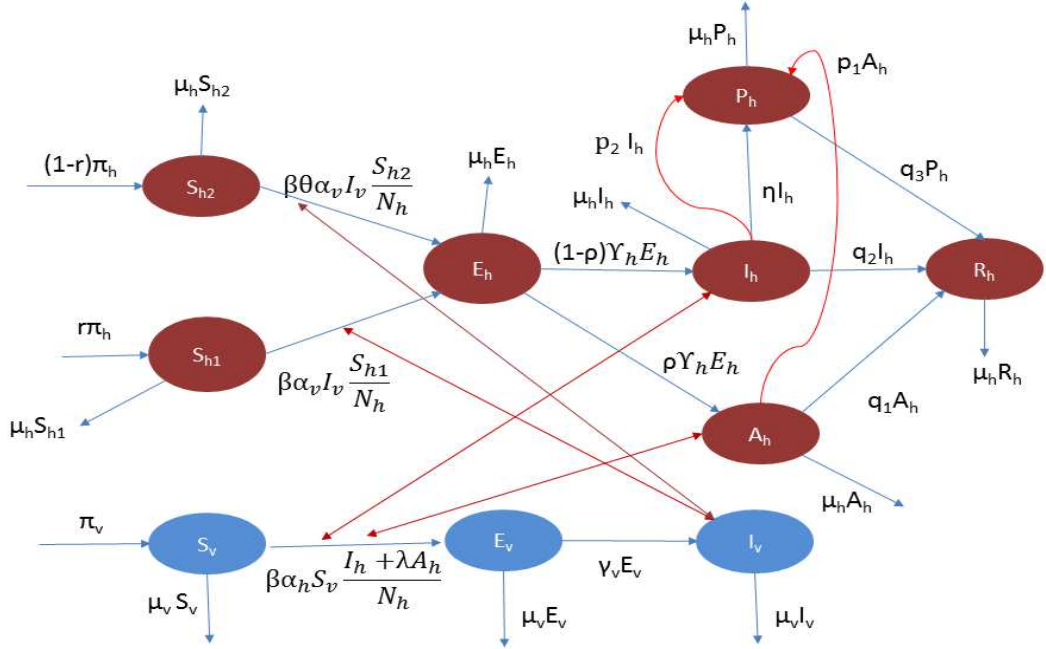}
   \caption{Schematic diagram of the model (\ref{eq1}). Two sided arrow represents new infection term, one sided arrow indicates progression to other compartments.}
   \label{flow_chart}
\end{figure}

The compartmental flow diagram is depicted in Fig. \ref{flow_chart}.
Keeping the above assumptions in mind, we developed the following
mathematical model for the transmission dynamics of dengue:
\begin{eqnarray}\label{eq1}
\frac{dS_v}{dt}&=&\pi_v-\beta\alpha_hS_v\left(\frac{I_h+\lambda A_h}{N_h}\right)-\mu_vS_v,\nonumber\\
\frac{dE_v}{dt}&=&\beta\alpha_hS_v\left(\frac{I_h+\lambda A_h}{N_h}\right)-(\gamma_v+\mu_v)E_v,\nonumber\\
\frac{dI_v}{dt}&=&\gamma_vE_v-\mu_vI_v,\nonumber\\
\frac{dS_{h1}}{dt}&=&r\pi_h-\beta\alpha_vI_v\left(\frac{S_{h1}}{N_h}\right)-\mu_hS_{h1},\nonumber\\
\frac{dS_{h2}}{dt}&=&(1-r)\pi_h-\beta\theta\alpha_vI_v\left(\frac{S_{h2}}{N_h}\right)-\mu_hS_{h2},\nonumber\\
\frac{dE_h}{dt}&=&\beta\alpha_vI_v\left(\frac{S_{h1}+\theta S_{h2}}{N_h}\right)-(\gamma_h+\mu_h)E_h,\\
\frac{dA_h}{dt}&=&\rho\gamma_hE_h-(\mu_h+q_1+p_1)A_h,\nonumber\\
\frac{dI_h}{dt}&=&(1-\rho)\gamma_hE_h-(\mu_h+\eta+q_2+p_2)I_h,\nonumber\\
\frac{dP_h}{dt}&=&p_1A_h+(\eta+p_2)I_h-(q_3+\mu_h+\delta)P_h,\nonumber\\
\frac{dR_h}{dt}&=&q_1A_h+q_2I_h+q_3P_h-\mu_hR_h.\nonumber
\end{eqnarray}

All parameters involved in the system (\ref{eq1}) are assumed to be
positive and also the initial conditions are taken to be positive
values. The biological meanings of variables and parameters involved
in the system (\ref{eq1}) are given in Tables \ref{table0} and
\ref{table1}, respectively.
\begin{center}
\begin{table}[!htbp]
\caption{Descriptions of variables used in the system (\ref{eq1})}
\begin{tabular}{|c|c|}
 \hline
Variables & Descriptions \\
\hline
 $S_v$ & Number of susceptible mosquito population \\
 $E_v$ & Number of exposed mosquito population \\
 $I_v$ & Number of infected mosquito population \\
 $S_{h1}$ & Number of high risk susceptible human population \\
 $S_{h2}$ & Number of low risk susceptible human population \\
 $E_h$ & Number of exposed human population \\
 $A_h$ & Number of asymptomatic human population \\
 $I_h$ & Number of symptomatic human population \\
 $P_h$ & Number of hospitalized and/or notified human population \\
 $R_h$ & Number of recovered human population \\
 \hline
\end{tabular}
\label{table0}
\end{table}
\end{center}
\begin{center}
\begin{table}[!htbp]
\caption{Biological meanings of parameters involved in the system
(\ref{eq1})}
\begin{tabular}{|c|c|c|c|c|}
 \hline
Parameters & Descriptions & Units & Values \\
\hline
 $\pi_v$ & Recruitment rate of adult susceptible mosquito population & year$^{-1}$ & 350000 \\
 $\beta$ & Average biting rate per mosquito per person & year$^{-1}$ & Estimated \\
 $\alpha_h$ & Transmission probability from infected human & --- & 0.75 \\
 & to susceptible mosquito & & \\
 $\lambda$ & Relative infectiousness of asymptomatic humans in relation & --- & 0.5  \\
 & to symptomatic humans & & \\
 $\mu_v$ & Natural death rate of adult mosquito population & year$^{-1}$ & 3 \\
 $\gamma_v$ & Intrinsic incubation & year$^{-1}$ & 3.795  \\
 $\pi_h$ & Recruitment rate of susceptible human & year$^{-1}$ & 273600 \\
 $r$ & Fraction of newly recruited individuals joining the high risk & --- & 0.25 \\
 & susceptible class & & \\
 $\alpha_v$ & Transmission probability from infected mosquito to & --- & 0.75 \\
 & susceptible human & & \\
 $\mu_h$ & Natural death rate of human & year$^{-1}$ & 0.0154 \\
 $\gamma_h$ & Extrinsic incubation & year$^{-1}$ & 3.3 \\
 $\theta$ & Relative chance of infection of low risk susceptible in relation & --- & 0.5 \\
 & to high risk susceptible & & \\
 $\rho$ & Fraction of exposed human population joining the & --- & 0.75 \\
 & asymptomatic class & & \\
 $p_1$ & Active case finding rate of asymptomatic class & year$^{-1}$ & Varied \\
 $p_2$ & Active case finding rate of symptomatic class & year$^{-1}$ & Varied \\
 $\eta$ & Rate of hospitalization and/or notification of symptomatic human & year$^{-1}$ & Estimated \\
 $q_1$ & Natural recovery rate of asymptomatic human & year$^{-1}$ & 4 \\
 $q_2$ & Natural recovery rate of symptomatic human & year$^{-1}$ & 0.0355 \\
 $q_3$ & Recovery rate of hospitalized and/or notified human & year$^{-1}$ & 4.5972 \\
 $\delta$ & Disease related death rate of human & year$^{-1}$ & 0.0001 \\ \hline
\end{tabular}
\label{table1}
\end{table}
\end{center}

It is worth noting that the feasible region for system (\ref{eq1})
is given in the following lemma \citep{Freedman1985,Hale1969}.
\begin{lemma}
The region of attraction for all solutions initiating in the
positive orthant is given by the set $\Omega$:
\begin{equation}\label{eqsv}
\Omega=\{(N_v,S_{h1},S_{h2},N_h): \ 0 \leq
N_v(t) \leq Z_1, \ 0
\leq S_{h1} \leq Z_2, \ 0
\leq S_{h2} \leq Z_3, \ Z_5 \leq N_h \leq
Z_4\},
\end{equation}
where
\begin{eqnarray*}
&&Z_1=\max\left\{\frac{\pi_v}{\mu_v},N_v(0)\right\}, \ Z_2=\max\left\{\frac{r\pi_h}{\mu_h},S_{h1}(0)\right\}, \ Z_3=\max\left\{\frac{(1-r)\pi_h}{\mu_h},S_{h2}(0)\right\},\\
&&Z_4=\max\left\{\frac{\pi_h}{\mu_h},N_h(0)\right\}, \ Z_5=\min\left\{\frac{\pi_h}{\mu_h+\delta},N_h(0)\right\},
\end{eqnarray*}
which is compact and invariant with respect to system (\ref{eq1}).
\end{lemma}

For proof of this lemma, see \textit{Appendix A}.

We first analyze the system through its equilibrium points.
Equilibrium points are the values of $S_v$, $E_v$, $I_v$, $S_{h1}$,
$S_{h2}$, $E_h$, $A_h$, $I_h$, $P_h$ and $R_h$ that remain constant
over time. The equilibrium points of system (\ref{eq1}) can be
obtained by equating the derivatives to zero.

\section{Disease-free equilibrium and its stability}\label{disease_free}

The disease-free equilibrium of the system (\ref{eq1}) is
$\displaystyle
E_0=\left(\frac{\pi_v}{\mu_v},0,0,\frac{r\pi_h}{\mu_h},\frac{(1-r)\pi_h}{\mu_h},0,0,0,0,0\right),$\
which is always feasible. Local stability of the equilibrium $E_0$
can be established in terms of the basic reproduction number
($\mathcal{R}_0$), a potential measure which determine that a
disease invade a population.

\subsection{Basic reproduction number}

Using next-generation operator method \cite{van2002}, we determine
the expression for basic reproduction number. For this, we find the
matrices $F$ (of new infection terms) and $V$ (of the transition
terms), as follows:

$F=\left(%
                  \begin{array}{cccccc}

                  0 & 0 & 0 & \displaystyle \frac{\beta\lambda\alpha_h\mu_h\pi_v}{\pi_h\mu_v} & \displaystyle \frac{\beta\alpha_h\mu_h\pi_v}{\pi_h\mu_v} & 0 \\
                  0 & 0 & 0 & 0 & 0 & 0 \\
                  0 & \beta\alpha_v[r+\theta(1-r)] & 0 & 0 & 0 & 0 \\
                  0 & 0 & 0 & 0 & 0 & 0 \\
                  0 & 0 & 0 & 0 & 0 & 0 \\
                  0 & 0 & 0 & 0 & 0 & 0 \\
                  \end{array}%
                    \right),$

$V=\left(%
                 \begin{array}{cccccc}

                  \gamma_v+\mu_v & 0 & 0 & 0 & 0 & 0 \\
                  -\gamma_v & \mu_v & 0 & 0 & 0 & 0 \\
                  0 & 0 & \gamma_h+\mu_h & 0 & 0 & 0 \\
                  0 & 0 & -\rho\gamma_h & \mu_h+q_1+p_1 & 0 & 0 \\
                  0 & 0 & -(1-\rho)\gamma_h & 0 & \mu_h+\eta+q_2+p_2 & 0 \\
                  0 & 0 & 0 & -p_1 & -(\eta+p_2) & q_3+\mu_h+\delta \\
                 \end{array}%
                 \right).$

The basic reproduction number is given by
$\mathcal{R}_0=\rho(FV^{-1})$, where $\rho$ is the spectral radius
of the next-generation matrix $(FV^{-1})$. Thus, from the model
(\ref{eq1}), we obtain the expression for $\mathcal{R}_0$ as
\begin{eqnarray}\label{eq_R0}
\mathcal{R}^2_0=\frac{\beta^2\pi_v\alpha_v\alpha_h\mu_h\gamma_v\gamma_h\{r+\theta(1-r)\}}{\mu^2_v\pi_h(\gamma_v+\mu_v)(\gamma_h+\mu_h)}
\left[\frac{\lambda\rho}{\mu_h+q_1+p_1}+\frac{1-\rho}{\eta+\mu_h+q_2+p_2}\right].
\end{eqnarray}

The quantity $\mathcal{R}_0$ is known as \lq\lq basic reproduction
number, the expected number of secondary cases produced in
completely susceptible population, by a typical infective
individual\rq\rq \ for the system (\ref{eq1}). From the expression
of $\mathcal{R}_0$, the role of active case finding for asymptomatic
as well as symptomatic individuals on the disease prevalence are
evident. It is to be noted that by increasing the rate of
hospitalization and/or notification of symptomatic humans and active case finding of
asymptomatic and symptomatic individuals, the values of
$\mathcal{R}_0$ decreases. The parameters $\beta$ and $\mu_v$ appear
in square terms and hence affects the values of $\mathcal{R}_0$
significantly. The former increases the values of $\mathcal{R}_0$
while the latter decreases the values of $\mathcal{R}_0$.

Following \citep{van2002}, regarding local stability of the
disease-free equilibrium $E_0$ of the system (\ref{eq1}), we have
the following theorem.

\begin{theorem}\label{thm_E0}
For system (\ref{eq1}), the disease-free equilibrium $E_0$ is
locally asymptotically stable if $\mathcal{R}_0<1$ and unstable if
$\mathcal{R}_0>1$.
\end{theorem}
For proof of this theorem, see \textit{Appendix B}.
\begin{remark}
The above theorem imply that whenever $\mathcal{R}_0$ is less than
unity, a small influx of infected mosquitoes/humans into the
community would not generate large outbreaks, and the disease dies
out in time.
\end{remark}

\section{Endemic equilibrium and its stability}\label{endemic}
\subsection{Existence of endemic equilibrium}

For model (\ref{eq1}), an endemic equilibrium is
$E^*=(S^*_v,E^*_v,I^*_v,S^*_{h1},S^*_{h2},A^*_h,E^*_h,I^*_h,P^*_h,R^*_h)$,
whose components are positive solutions of equilibrium equations of
the system (\ref{eq1}).

We define a new variable $Q$ as
\begin{eqnarray}\label{eq2a}
Q=\beta\alpha_v\frac{I^*_v}{N^*_h}.
\end{eqnarray}

From the equilibrium equations of system (\ref{eq1}), we have
\begin{align}\label{eq2b}
S^*_v &= \frac{\pi_v}{\left[\mu_v+\beta\alpha_h\left(\frac{I^*_h+\lambda A^*_h}{N^*_h}\right)\right]}, & E^*_v &=\frac{\beta\alpha_h\pi_v\left(\frac{I^*_h+\lambda A^*_h}{N^*_h}\right)}{(\gamma_v+\mu_v)\left[\mu_v+\beta\alpha_h\left(\frac{I^*_h+\lambda
A^*_h}{N^*_h}\right)\right]},\nonumber\\
I^*_v &=\frac{\beta\alpha_h\gamma_v\pi_v\left(\frac{I^*_h+\lambda
A^*_h}{N^*_h}\right)}{\mu_v(\gamma_v+\mu_v)\left[\mu_v+\beta\alpha_h\left(\frac{I^*_h+\lambda
A^*_h}{N^*_h}\right)\right]}, & S^*_{h1} &=\frac{r\pi_h}{Q+\mu_h},\nonumber\\
S^*_{h2} &=\frac{(1-r)\pi_h}{\theta Q+\mu_h}, & E^*_h &=\frac{Q}{\mu_h+\gamma_h}\left[\frac{r\pi_h}{Q+\mu_h}+\theta\frac{(1-r)\pi_h}{\theta
Q+\mu_h}\right],\nonumber\\
A^*_h &=\frac{\rho\gamma_hE^*_h}{\mu_h+q_1+p_1}, & I^*_h &=\frac{(1-\rho)\gamma_hE^*_h}{\mu_h+\eta+q_2+p_2}, \nonumber\\
P^*_h
&=\frac{\gamma_hE^*_h}{\mu_h+\delta+q_3}\left[\frac{p_1\rho}{\mu_h+q_1+p_1}+\frac{(1-\rho)(\eta+p_2)}{\mu_h+\eta+q_2+p_2}\right],
& R^*_h &=\frac{q_1A^*_h+q_2I^*_h+q_3P^*_h}{\mu_h}.
\end{align}

The total human population is given by
\begin{eqnarray}\label{eq2l}
N^*_h=\frac{1}{\mu_h}\left[\pi_h-\frac{\delta\gamma_hE^*_h}{\mu_h+\delta+q_3}\left\{\frac{p_1\rho}{\mu_h+q_1+p_1}+\frac{(1-\rho)(\eta+p_2)}{\mu_h+\eta+q_2+p_2}\right\}\right].
\end{eqnarray}

Now, using equations (\ref{eq2b}) and (\ref{eq2l}) in equation (\ref{eq2a}), we get the following equation in $Q$:
\begin{eqnarray}\label{eq_endemic}
C_4Q^4+C_3Q^3+C_2Q^2+C_1Q+C_0=0,
\end{eqnarray}
where
\begin{eqnarray*}
&&C_4=\theta^2(X_2-X_3)(X_5+X_6),\\
&&C_3=\mu_h\theta(X_2-X_3)[X_5(1+\theta)+X_6\{r+\theta(1-r)\}]+\theta\mu_h(X_5+X_6)[X_2(1+\theta)-X_3\{r+\theta(1-r)\}]\\
&&\hskip 1cm -X_1X_4\theta^2,\\
&&C_2=\theta X_5\mu^2_h(X_2-X_3)+\mu^2_h[X_2(1+\theta)-X_3\{r+\theta(1-r)\}][X_5(1+\theta)+X_6\{r+\theta(1-r)\}]\\
&&\hskip 1cm +X_2\theta\mu^2_h(X_5+X_6)-X_1X_4\theta\mu_h\{r+\theta(1-r)\}-X_1X_4\mu_h\theta(1+\theta),\\
&&C_1=X_5\mu^3_h[X_2(1+\theta)-X_3\{r+\theta(1-r)\}]+X_2\mu^3_h[X_5(1+\theta)+X_6\{r+\theta(1-r)\}]\\
&&\hskip 1cm -X_1X_4\mu^2_h(1+\theta)\{r+\theta(1-r)\}-X_1X_4\theta\mu^2_h,\\
&&C_0=X_2X_5\mu^4_h-X_1X_4\mu^3_h\{r+\theta(1-r)\}=\pi^2_h\mu^2_h\mu_v(1-R^2_0)
\end{eqnarray*}
with
\begin{eqnarray*}
&&X_1=\frac{\pi_h\gamma_h}{\mu_h+\gamma_h}\left[\frac{1-\rho}{\mu_h+\eta+q_2+p_2}+\frac{\lambda\rho}{\mu_h+q_1+p_1}\right],
\ X_2=\frac{\pi_h}{\mu_h},\\
&&X_3=\frac{\delta\pi_h\gamma_h}{\mu_h(\mu_h+\gamma_h)(\mu_h+\delta+q_3)}\left[\frac{p_1\rho}{\mu_h+q_1+p_1}+\frac{(1-\rho)(\eta+p_2)}{\mu_h+\eta+q_2+p_2}\right],\\
&&X_4=\frac{\beta^2\alpha_v\alpha_h\pi_v\gamma_v}{\mu_v(\mu_v+\gamma_v)},
\ X_5=\mu_vX_2, \ X_6=\beta\alpha_hX_1-\mu_vX_3.
\end{eqnarray*}

By employing the Descartes' rule of signs on the equation given in
(\ref{eq_endemic}), we list the various possibilities for the
positive roots of this equation in Table \ref{table_roots}
\cite{Wang2004}.

{\tiny
\begin{center}
\begin{table}[!htbp]
\caption{Number of possible positive real roots of Eq.
(\ref{eq_endemic}) for $\mathcal{R}_0<1$ and $\mathcal{R}_0>1$.}
\begin{tabular}{ccccccccc}
 \hline
Cases & $C_4$ & $C_3$ & $C_2$ & $C_1$ & $C_0$ & $\mathcal{R}_0$ & No. of sign & No. of possible \\
& & & & & & & changes & positive real roots \\
\hline
 1. & + & + & + & + & + & $\mathcal{R}_0<1$ & 0 & 0 \\
    & + & + & + & + & - & $\mathcal{R}_0>1$ & 1 & 1 \\
 2. & + & - & - & - & + & $\mathcal{R}_0<1$ & 2 & 0,2 \\
    & + & - & - & - & - & $\mathcal{R}_0>1$ & 1 & 1 \\
 3. & + & - & - & + & + & $\mathcal{R}_0<1$ & 2 & 0,2 \\
    & + & + & - & - & - & $\mathcal{R}_0>1$ & 1 & 1 \\
 4. & + & - & + & - & + & $\mathcal{R}_0<1$ & 4 & 0,2,4 \\
    & + & - & + & - & - & $\mathcal{R}_0>1$ & 3 & 1,3 \\
 5. & + & - & - & + & + & $\mathcal{R}_0<1$ & 2 & 0,2 \\
    & + & - & - & + & - & $\mathcal{R}_0>1$ & 3 & 1,3 \\
 6. & + & + & + & - & + & $\mathcal{R}_0<1$ & 2 & 0,2 \\
    & + & + & + & - & - & $\mathcal{R}_0>1$ & 1 & 1 \\
 7. & + & + & - & + & + & $\mathcal{R}_0<1$ & 2 & 0,2 \\
    & + & + & - & + & - & $\mathcal{R}_0>1$ & 3 & 1,3 \\
 8. & + & - & + & + & + & $\mathcal{R}_0<1$ & 2 & 0,2 \\
    & + & - & + & + & - & $\mathcal{R}_0>1$ & 3 & 1,3 \\
 9. & - & + & + & + & + & $\mathcal{R}_0<1$ & 1 & 1 \\
    & - & + & + & + & - & $\mathcal{R}_0>1$ & 2 & 0,2 \\
 10. & - & - & - & - & + & $\mathcal{R}_0<1$ & 1 & 1 \\
    & - & - & - & - & - & $\mathcal{R}_0>1$ & 0 & 0 \\
 11. & - & - & - & + & + & $\mathcal{R}_0<1$ & 1 & 1 \\
    & - & + & - & - & - & $\mathcal{R}_0>1$ & 2 & 0,2 \\
 12. & - & - & + & - & + & $\mathcal{R}_0<1$ & 3 & 1,3 \\
    & - & - & + & - & - & $\mathcal{R}_0>1$ & 2 & 0,2 \\
 13. & - & - & - & + & + & $\mathcal{R}_0<1$ & 1 & 1 \\
    & - & - & - & + & - & $\mathcal{R}_0>1$ & 2 & 0,2 \\
 14. & - & + & + & - & + & $\mathcal{R}_0<1$ & 3 & 1,3 \\
    & - & + & + & - & - & $\mathcal{R}_0>1$ & 2 & 0,2 \\
 15. & - & + & - & + & + & $\mathcal{R}_0<1$ & 3 & 1,3 \\
    & - & + & - & + & - & $\mathcal{R}_0>1$ & 4 & 0,2,4 \\
 16. & - & - & + & + & + & $\mathcal{R}_0<1$ & 1 & 1 \\
    & - & - & + & + & - & $\mathcal{R}_0>1$ & 2 & 0,2 \\
    \hline
\end{tabular}
\label{table_roots}
\end{table}
\end{center}}

\subsubsection{Global stability of the endemic equilibrium}

Using the fact that $N_h=S_{h1}+S_{h2}+E_h+A_h+I_h+P_h+R_h$, we have the following system:
\begin{eqnarray}\label{eqg1}
\frac{dS_v}{dt}&=&\pi_v-\beta\alpha_hS_v\left(\frac{I_h+\lambda A_h}{N_h}\right)-\mu_vS_v,\nonumber\\
\frac{dE_v}{dt}&=&\beta\alpha_hS_v\left(\frac{I_h+\lambda A_h}{N_h}\right)-(\gamma_v+\mu_v)E_v,\nonumber\\
\frac{dI_v}{dt}&=&\gamma_vE_v-\mu_vI_v,\nonumber\\
\frac{dN_h}{dt}&=&\pi_h-\mu_hN_h-\delta P_h,\nonumber\\
\frac{dS_{h2}}{dt}&=&(1-r)\pi_h-\beta\theta\alpha_vI_v\left(\frac{S_{h2}}{N_h}\right)-\mu_hS_{h2},\nonumber\\
\frac{dE_h}{dt}&=&\beta\alpha_vI_v\left(\frac{(N_h-S_{h2}-E_h-A_h-I_h-P_h-R_h)+\theta S_{h2}}{N_h}\right)-(\gamma_h+\mu_h)E_h,\\
\frac{dA_h}{dt}&=&\rho\gamma_hE_h-(\mu_h+q_1+p_1)A_h,\nonumber\\
\frac{dI_h}{dt}&=&(1-\rho)\gamma_hE_h-(\mu_h+\eta+q_2+p_2)I_h,\nonumber\\
\frac{dP_h}{dt}&=&p_1A_h+(\eta+p_2)I_h-(q_3+\mu_h+\delta)P_h,\nonumber\\
\frac{dR_h}{dt}&=&q_1A_h+q_2I_h+q_3P_h-\mu_hR_h.\nonumber
\end{eqnarray}

Since system (\ref{eqg1}) is equivalent to the system (\ref{eq1}), we study the global asymptotic stability of the endemic equilibrium
$E^*(S^*_v,E^*_v,I^*_v,N^*_h,S^*_{h2},E^*_h,A^*_h,I^*_h,P^*_h,R^*_h)$ of the system (\ref{eqg1}).

Regarding global asymptotic stability of the equilibrium $E^*$, we have the following theorem.

\begin{theorem}\label{globalstability_E*}
The equilibrium $E^*$ is globally asymptotically stable inside the region of attraction $\Omega$, provided the following inequalities hold:
{\small
\begin{eqnarray}
&&\max\left\{\frac{15}{2\mu_h}\left[\frac{\beta\alpha_v\{I^*_v(S^*_{h2}+E^*_h+A^*_h+I^*_h+P^*_h+R^*_h)+\theta\pi_h\pi_v/(\mu_v\mu_h)\}}
{N^*_h\pi_h/(\mu_h+\delta)}\right]^2,\right.\nonumber\\
&&\left.\hskip 1cm \frac{9}{2\mu_v}\left[\beta\alpha_v\left\{\frac{\pi_h/\mu_h-S^*_{h2}-E^*_h-A^*_h-I^*_h-P^*_h-R^*_h}{\pi_h/(\mu_h+\delta)}
+\frac{\theta\pi_h}{\mu_hN^*_h}\right\}\right]^2,\right.\nonumber\\
&&\left.\hskip 1cm \frac{9N^*_h}{2(\beta\theta\alpha_vI^*_v+\mu_hN^*_h)}\left[\frac{\beta\theta\alpha_vI^*_v}{N^*_h}\right]^2,
\frac{15}{2(\mu_h+\delta+q_3)}\left[\frac{\beta\alpha_v\pi_v(\mu_h+\delta)}{\mu_v\pi_h}\right]^2,\right.\nonumber\\
&&\left. \hskip 1cm
\frac{6}{\mu_h}\left[\frac{\beta\alpha_v\pi_v(\mu_h+\delta)}{\mu_v\pi_h}\right]^2,
\frac{15[\rho\gamma_h]^2}{2(\mu_h+q_1+p_1)},\frac{15[(1-\rho)\gamma_h]^2}{2(\mu_h+\eta+q_2+p_2)}\right\}<[\gamma_h+\mu_h],\label{eqgl3}
\end{eqnarray}
\begin{eqnarray}
&&\max\left\{\frac{5}{\mu_h}\left[\frac{\beta\alpha_h\pi_v(1+\lambda)(\mu_h+\delta)}{\mu_v\mu_hN^*_h}\right]^2,
\frac{5}{\mu_h+\eta+q_2+p_2}\left[\frac{\beta\alpha_h\pi_v}{\mu_v\mu_hN^*_h}\right]^2,
\frac{5}{\mu_h+q_1+p_1}\left[\frac{\beta\lambda\alpha_h\pi_v}{\mu_vN^*_h}\right]^2,\frac{3\gamma^2}{\mu_v}\right\}\nonumber\\
&&\hskip 3cm<[\gamma_v+\mu_v],\label{eqgl4}
\end{eqnarray}
\begin{eqnarray}
&&\max\left\{\frac{5}{\mu_h}\left[\frac{\beta\alpha_h\pi_v(1+\lambda)(\mu_h+\delta)}{\mu_v\mu_hN^*_h}\right]^2,
\frac{5}{\mu_h+\eta+q_2+p_2}\left[\frac{\beta\alpha_h\pi_v}{\mu_vN^*_h}\right]^2,\frac{5}{\mu_h+q_1+p_1}\left[\frac{\beta\lambda\alpha_h}{N^*_h}\right]^2,\right.\nonumber\\
&&\left.\hskip 2cm
\frac{4}{\gamma_v+\mu_v}\left[\frac{\beta\alpha_h(I^*_h+\lambda
A^*_h)}{N^*_h}\right]^2\right\}<\left[\mu_v+\frac{\beta\alpha_h(I^*_h+\lambda
A^*_h)}{N^*_h}\right],\label{eqgl5}
\end{eqnarray}
\begin{eqnarray}
&&\max\left\{\frac{5q^2_1}{\mu_h+q_1+p_1},\frac{5q^2_2}{\mu_h+\eta+q_2+p_2},\frac{5q^2_3}{\mu_h+\delta+q_3},\frac{5\delta^2}{\mu_h+\delta+q_3},\right.\nonumber\\
&&\left.\hskip 3cm
\frac{15N^*_h}{4(\beta\theta\alpha_vI^*_v+\mu_hN^*_h)}\left[\frac{\beta\theta\alpha_v\pi_v(\mu_h+\delta)}{\mu_v\mu_hN^*_h}\right]^2\right\}<\mu_h,\label{eqgl6}
\end{eqnarray}
\begin{eqnarray}
\left[\frac{\beta\theta\pi_h\alpha_v}{\mu_hN^*_h}\right]^2<\frac{4}{9}\mu_v\left[\mu_h+\frac{\beta\theta\alpha_vI^*_v}{N^*_h}\right],\label{eqgl7}
\end{eqnarray}
\begin{eqnarray}
\max\left\{\frac{(\eta+p_2)^2}{\mu_h+\eta+q_2+p_2},\frac{p^2_1}{\mu_h+q_1+p_1}\right\}<\frac{4(\mu_h+\delta+q_3)}{25}.\label{eqgl8}
\end{eqnarray}}
\end{theorem}

For proof of this theorem, see \textit{Appendix C}.

\begin{remark}
Conditions of Theorem \ref{globalstability_E*} are only sufficient
for the global asymptotic stability of the equilibrium $E^*$ and
prevents persistent oscillations of the system solutions.
\end{remark}

To verify above theorem numerically, we choose the following set of hypothetical parameter values in the system (\ref{eq1})
\begin{eqnarray}\label{eqg19}
&&\pi_v=8, \ \beta=2, \ \mu_v=0.3, \ \gamma_v=0.795, \ \pi_h=0.15, \ \mu_h=0.154, \ \gamma_h=0.3,\nonumber\\
&&q_1=0.4, \ \delta=0.5, \ \alpha_h=0.75, \ \lambda=0.5, \ r=0.55, \
\alpha_v=0.05, \ \theta=0.85,\nonumber\\
&&\rho=0.25, \ q_2=0.355, \ q_3=0.5972, \ \eta=1.474\ p_1=0.01, \
p_2=0.02.
\end{eqnarray}
The components of the equilibrium $E^*$ are found to be
\begin{eqnarray*}
&&S^*_v=21.3803, \ E^*_v=1.4483, \ I^*_v=3.8380, \ S^*_{h1}=0.1385, \ S^*_{h2}=0.1275,\\
&&E^*_h=0.2402, \ A^*_h=0.0319, \ I^*_h=0.0270, \ P^*_h=0.0325, \ R^*_h=0.2711.
\end{eqnarray*}

\begin{figure}[t]
\includegraphics[width=.5\textwidth]{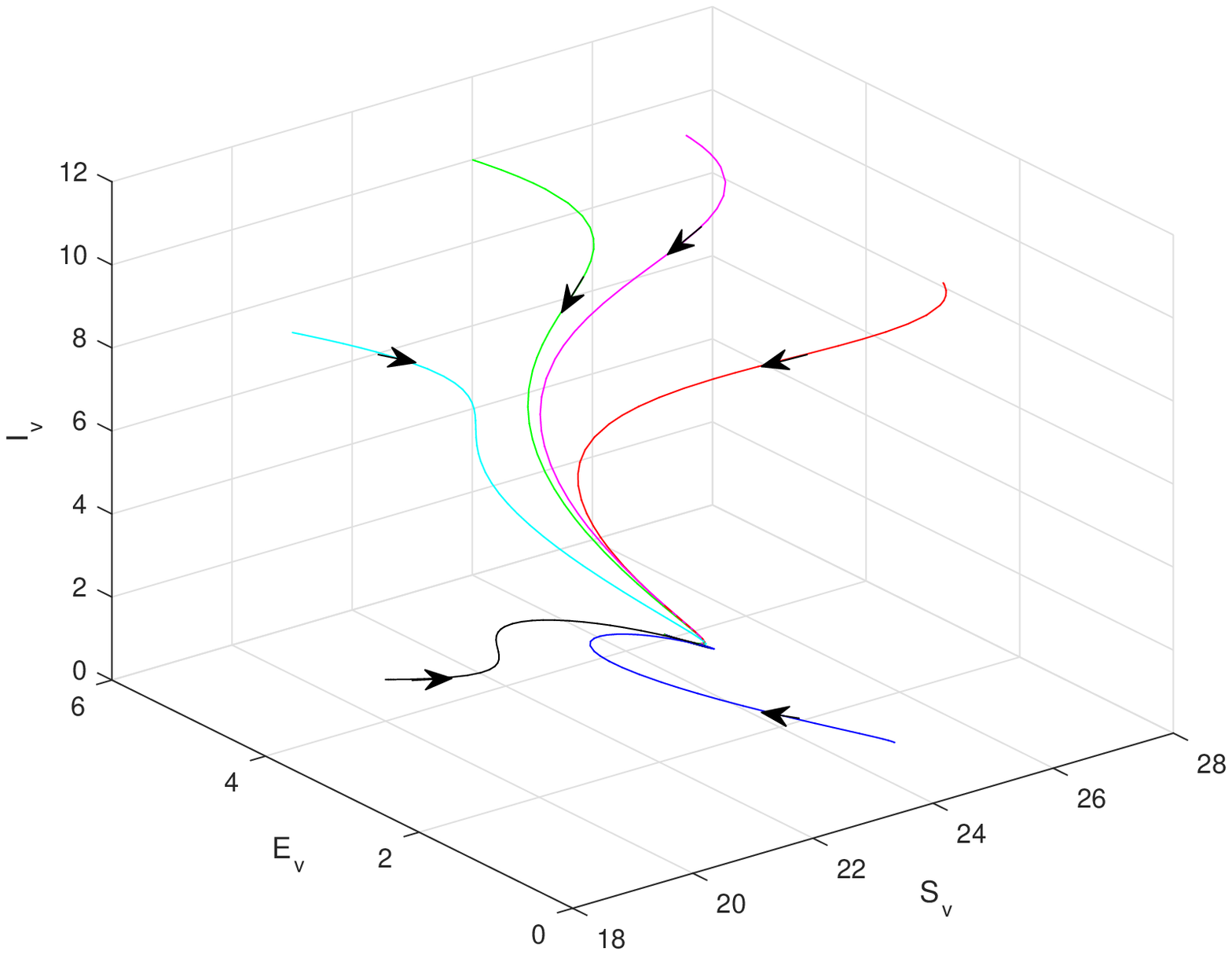}{a}
\includegraphics[width=.5\textwidth]{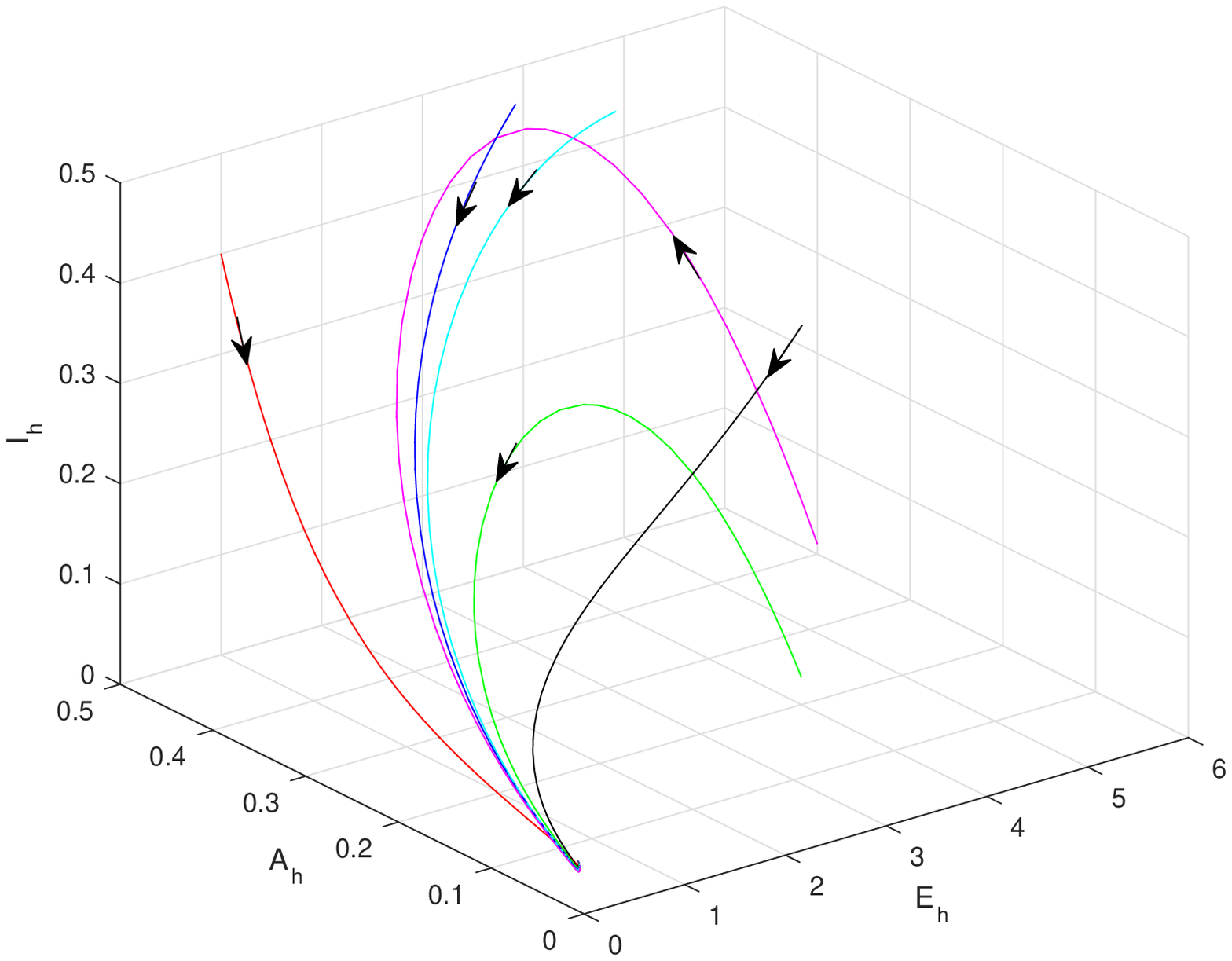}{b}
\caption{Global stability of the endemic equilibrium $E^*$ inside the region of attraction $\Omega$ in (a) $S_v$$-$$E_v$$-$$I_v$ and (b) $E_h$$-$$A_h$$-$$I_h$ spaces.}
\label{globalE*}
\end{figure}

For above set of parameter values, the conditions for the global
asymptotical stability of the equilibrium $E^*$ are satisfied. We
show the global stability of the endemic equilibrium $E^*$ inside
the region $\Omega$ in $S_v$$-$$E_v$$-$$I_v$ and
$E_h$$-$$A_h$$-$$I_h$ spaces, Fig. \ref{globalE*}. It is evident
from the figure that all the solution trajectories that originate
inside the region of attraction $\Omega$ approach the point
$(S^*_v,E^*_v,I^*_v)$ and $(E^*_h,A^*_h,I^*_h)$, as shown in Fig.
\ref{globalE*}a and Fig. \ref{globalE*}b, respectively. Thus, the
numerical results also confirm that the endemic equilibrium $E^*$ is
globally asymptotically stable in the $S_v$$-$$E_v$$-$$I_v$ and
$E_h$$-$$A_h$$-$$I_h$ spaces. Using this approach, we can show the
global asymptotic stability of the endemic equilibrium $E^*$ in
other spaces.

\section{Sensitivity analysis}\label{sensitivity1}
To see the effect of some controllable parameters, $\beta$, $\mu_v$,
$\eta$, $p_1$ and $p_2$, of the system (\ref{eq1}) on the value of
basic reproduction number, $\mathcal{R}_0$, we calculate the
normalized forward sensitivity indices of $\mathcal{R}_0$ to these
parameters. We evalauate the sensitivity indices for $\beta=42.885$,
$\eta=0.0119$, $p_1=0.3$ and $p_2=0.2$, and taking rest of the
parameter values from Table \ref{table1}. The normalized forward
sensitivity index for a variable $w$, which depends differentiably
on a parameter $\alpha$, is defined as
\begin{eqnarray*}
X^\alpha_w=\frac{\partial w}{\partial \alpha}\times
\frac{\alpha}{w}.
\end{eqnarray*}

The sensitivity indices of $\mathcal{R}_0$ with respect to the
parameters $\beta$, $\mu_v$, $\eta$, $p_1$ and $p_2$ are found to be
\begin{eqnarray*}
X^{\beta}_{\mathcal{R}_0}=1, \ X^{\mu_v}_{\mathcal{R}_0}=-1, \
X^{\eta}_{\mathcal{R}_0}=-0.1476, \ X^{p_1}_{\mathcal{R}_0}=-0.0255,
\ X^{p_2}_{\mathcal{R}_0}=-0.1990.
\end{eqnarray*}

The fact that $X^\beta_{\mathcal{R}_0}=1$ means that 1\% increase in
$\beta$, keeping other parameters fixed, will produce 1\% increase
in $\mathcal{R}_0$. When the parameters $\mu_v$, $\eta$, $p_1$ and
$p_2$ increase by $1$\% while keeping other parameters constant, the
value of $\mathcal{R}_0$ decreases by $1$\%, $0.1476$\%, $0.0255$\%
and $0.1990$\%, respectively. Overall a lower value of
$\mathcal{R}_0$ is preferable because it increases the possibility
of disease eradication in the region. Therefore, above all
prevention practices must focus on a decrease in the parameter
$\beta$, while an increase in the parameters $\mu_v$, $\eta$, $p_1$
and $p_2$ should instead be favored.

\begin{figure}[t]
\includegraphics[width=1\textwidth]{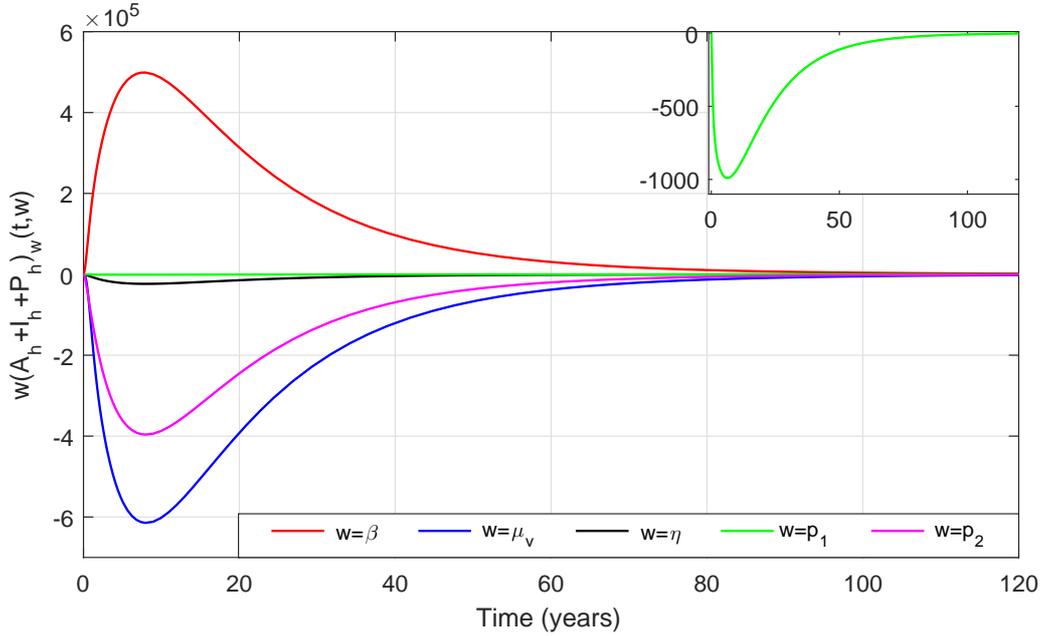}
\caption{semi-relative sensitivity solutions of total infective
population ($A_h+I_h+P_h$) with respect to $\beta$, $\mu_v$, $\eta$,
$p_1$ and $p_2$.} \label{sensitivity}
\end{figure}

Further, to check how the total infective populations
($A_h+I_h+P_h$) are affected with variations in average biting rate
of mosquito, death rate of mosquito, rate of hospitalization and/or notification of
symptomatic individuals, and ACF of asymptomatic and symptomatic
individuals, we perform semi-relative sensitivity analysis of the
system (\ref{eq1}) \cite{Bortz2004}. We plot the semi-relative
sensitivity solutions of the total infective populations
($A_h+I_h+P_h$) with respect to $\beta$, $\mu_v$, $\eta$. $p_1$ and
$p_2$ in Fig. \ref{sensitivity}. From the figure, we can see that
doubling of these parameters exhibit their largest influences early
in the simulation and a large expected variation in the total
infected populations is observed. It is apparent from the figure
that the doubling of $\beta$ and $\mu_v$ will yield sudden increase
and decrease of total infected populations, respectively around
$t=10$ years. Sudden increase in the total infected population on
doubling of $\beta$ is due to a large number of initial susceptible
population. As time progresses, the susceptible population decreases
and on natural recovery, the infective decreases. Similarly, the
infective decreases by a large number by doubling $\mu_v$ on its
initial phase but with increase in time, the infective first
increases and then again decreases. On the other hand, $\eta$
exhibits maximum reduction in total infected population around
$t=10$ years and no further reduction is observed. Such decrease in
total infected population is due to insufficient mosquito control
and personal protection. Increase in the parameters due to active
case finding reduce the prevalence of the disease. Note here that
the parameters $\beta$, $\mu_v$ and $p_2$ have larger effects in
comparison to the other two parameters on the total infective
population. Therefore, they play crucial roles for the control of
the disease. Moreover, the impacts of $\eta$ and $p_1$ are also
important.

\section{Data and model calibration}\label{estimation}
India experienced high levels of dengue cases in last few years. In
2017, a provisional total of 1,88,401 cases has been reported to
NVBDCP \cite{NVBDCP2018}. We use annual reported cases of dengue
fever in eight states of India to calibrate the model (\ref{eq1}) in
the absence of ACF parameters, $p_1$ and $p_2$. The reason behind
dropping the ACF parameters is that currently ACF is not employed in
India for dengue control. For our study, we choose Kerala, Delhi,
Gujarat, West Bengal, Andhra Pradesh, Rajasthan, Maharashtra and
Karnataka, dengue endemic states of India.

We estimate the unknown parameters $\widehat{\theta}=(\beta,\eta)$
using the annual new dengue cases from 2007 to 2017
\cite{Chakravarti2012,NVBDCP2018}. All the fixed parameters are
taken from Table \ref{table1}. Let $P(t,\widehat{\theta})$ denote the
number of new hospitalized and/or notified dengue cases from model (\ref{eq1}) at the $t^{th}$
year, then $P(t,\widehat{\theta})$ has the form
\begin{equation}
P(t,\widehat{\theta})=\int\limits_{t-1}^{t}\left[\eta I_h\right]dt,
\end{equation}
and if $C(0)$ is the number of new hospitalized and/or notified dengue cases at the first time point of
the data, then $P(0)=C(0)$. We have $R$ independent observations
from data, representing the number of new hospitalized and/or notified dengue cases in the $i^{th}$
year, where $i=1, \cdots, R$. Let $\epsilon$ be the error of fit,
which follows the Gaussian distribution having an unknown variance $\sigma^2$
\begin{eqnarray*}
Y_i=P(t_i,\widehat{\theta})+\epsilon, \epsilon\sim N(0,I\sigma^2).
\end{eqnarray*}

\begin{figure}[t]
\includegraphics[width=0.5\textwidth]{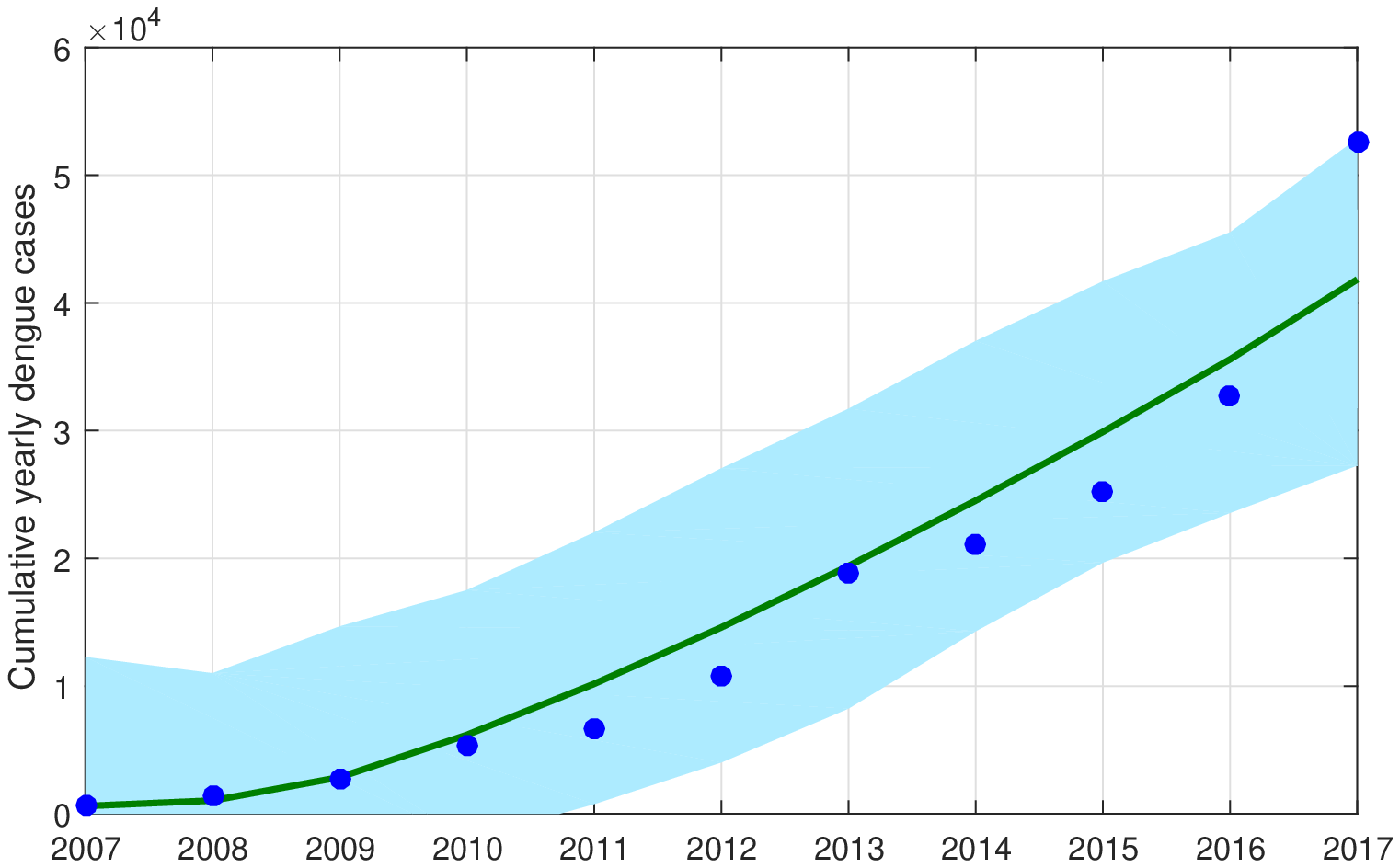}{a}
\includegraphics[width=0.5\textwidth]{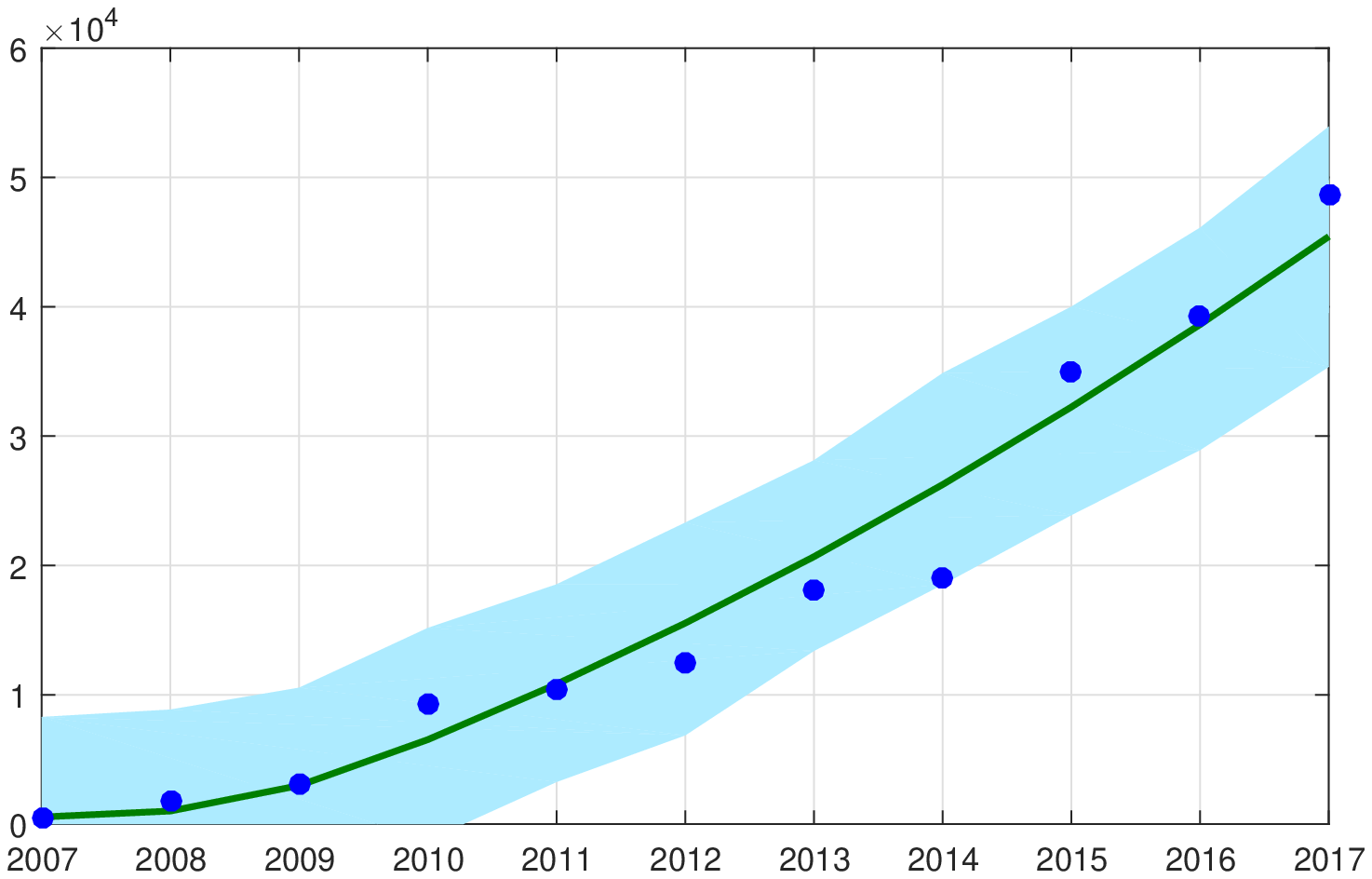}{b}
\includegraphics[width=0.5\textwidth]{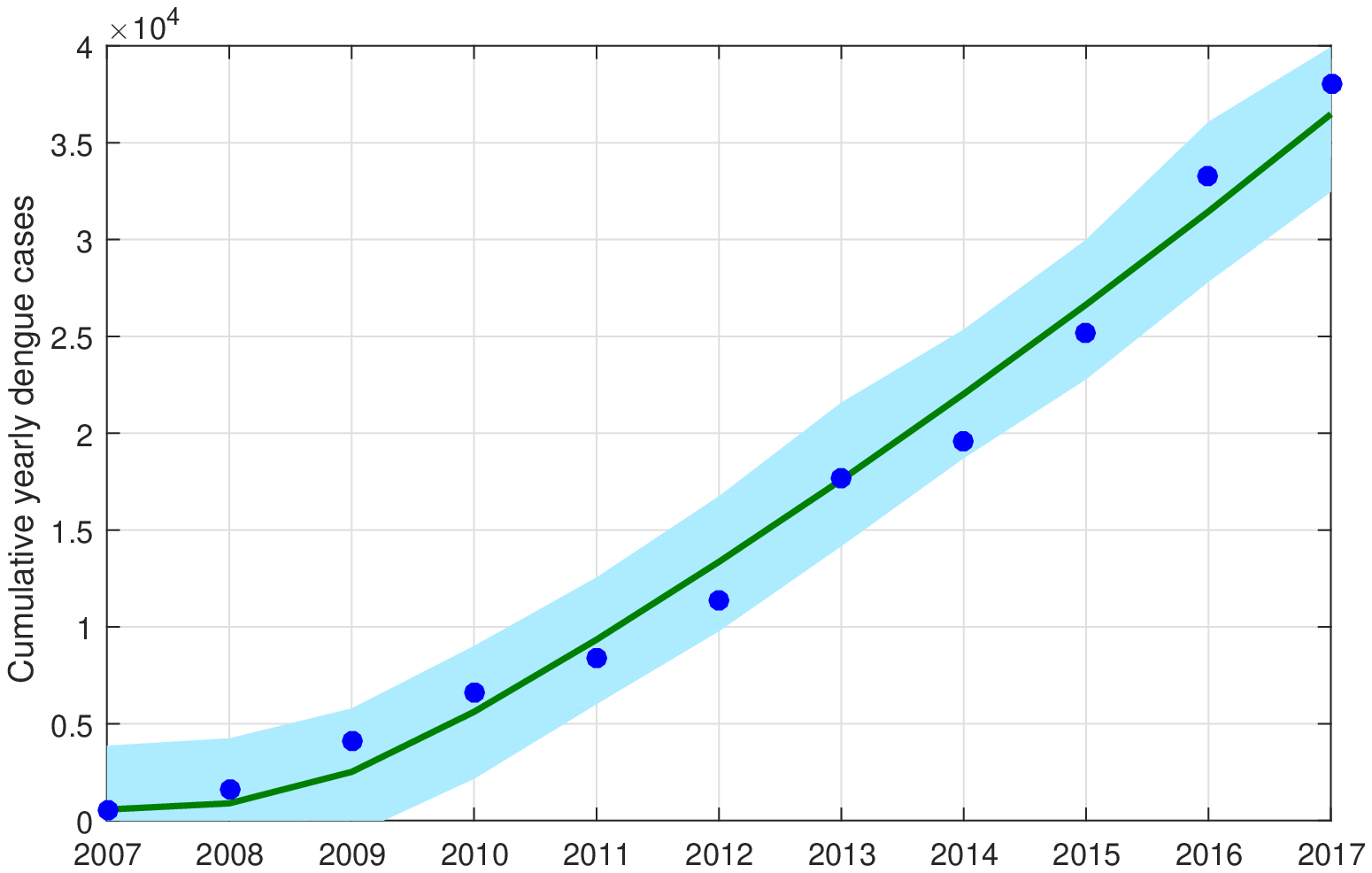}{c}
\includegraphics[width=0.5\textwidth]{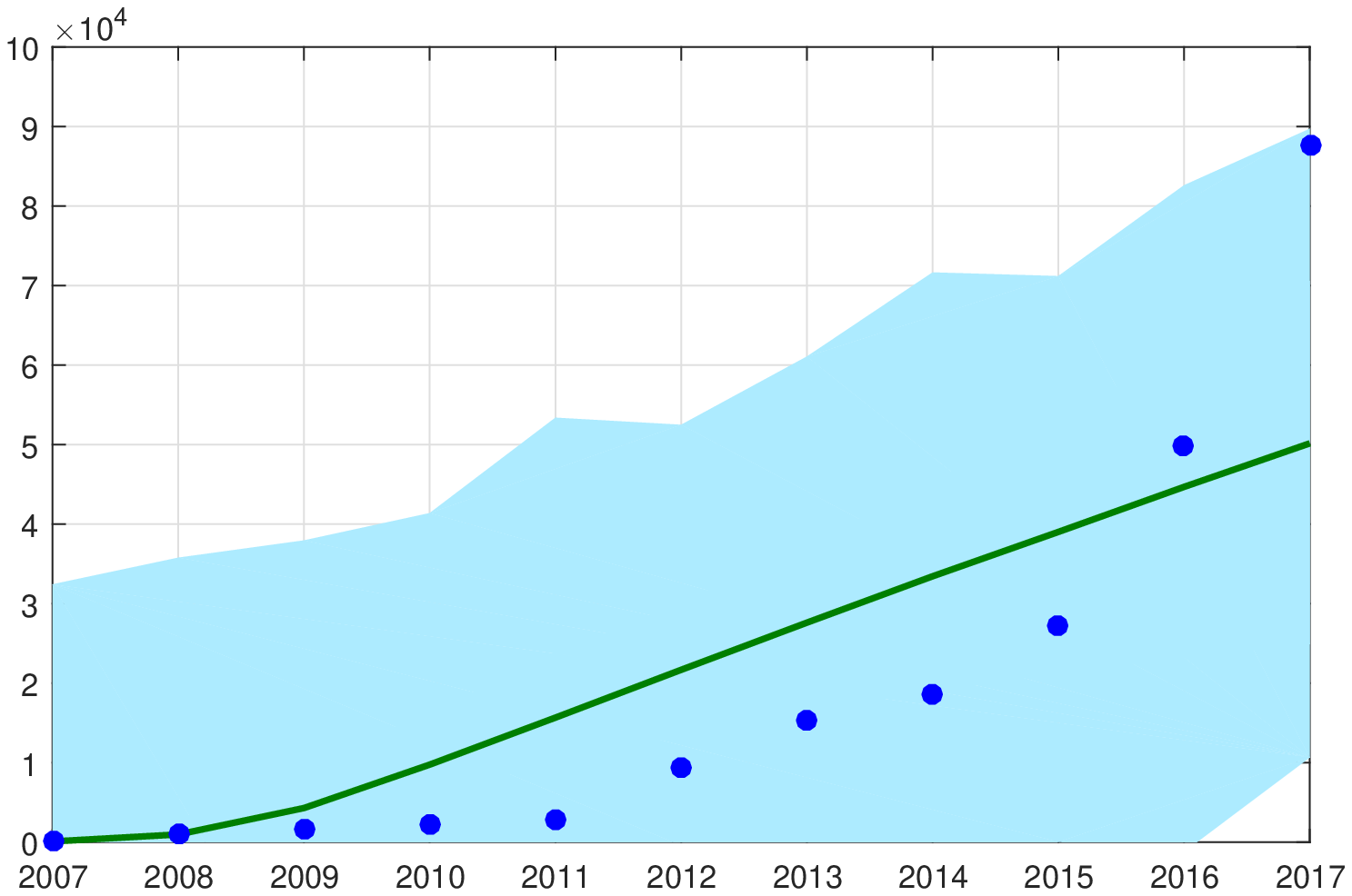}{d}
\caption{Plots of the output of the fitted model (\ref{eq1}) and the
observed cumulative dengue data for (a) Kerala, (b) Delhi, (c)
Gujarat and (d) West Bengal. Cumulative cases (filled blue circle)
from the data, and model simulated data (thick green curve) are
plotted with the parameter estimates using parameter values of Table
\ref{table1}.} \label{fig_fitting1}
\end{figure}

\begin{figure}[t]
\includegraphics[width=0.5\textwidth]{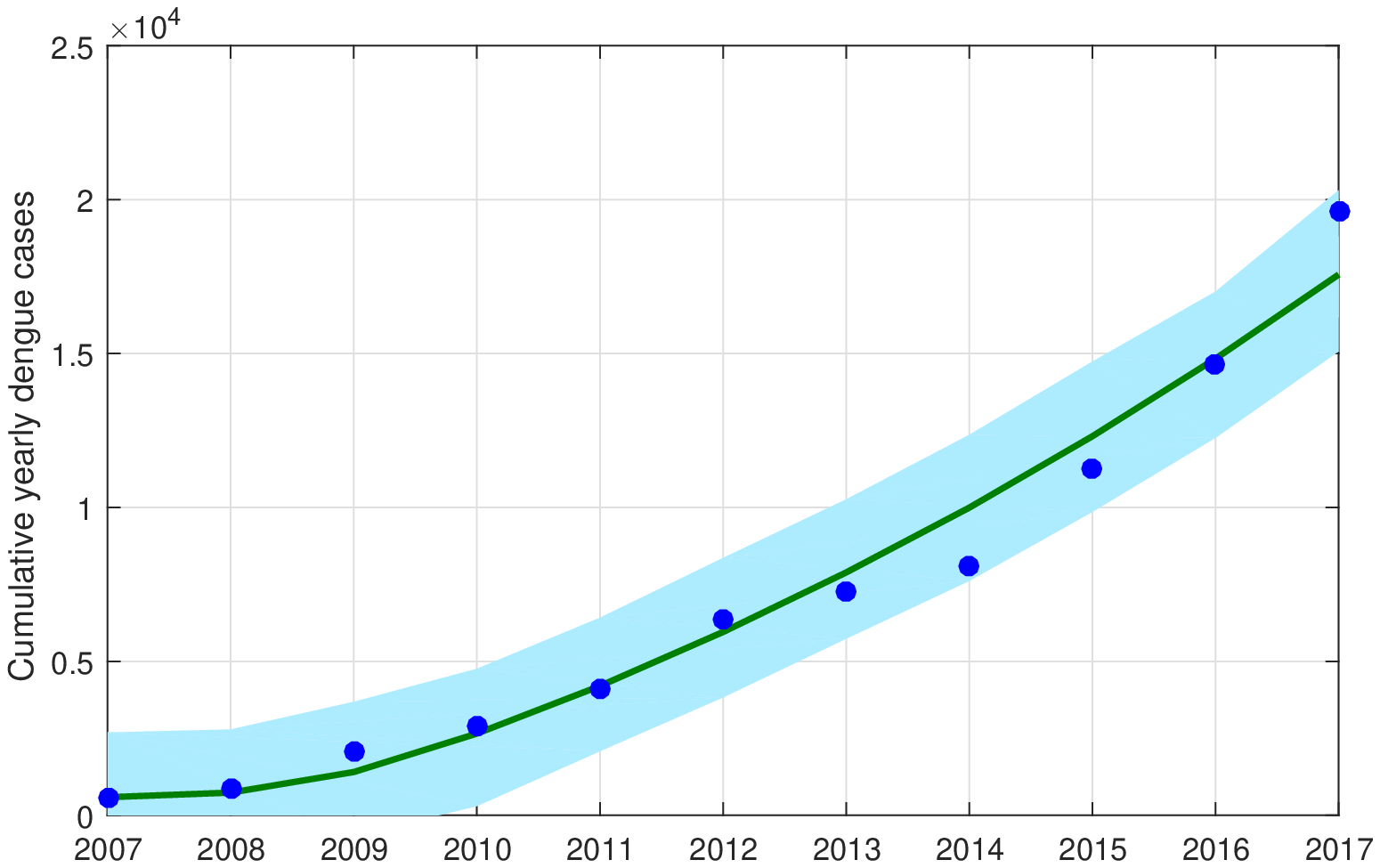}{a}
\includegraphics[width=0.5\textwidth]{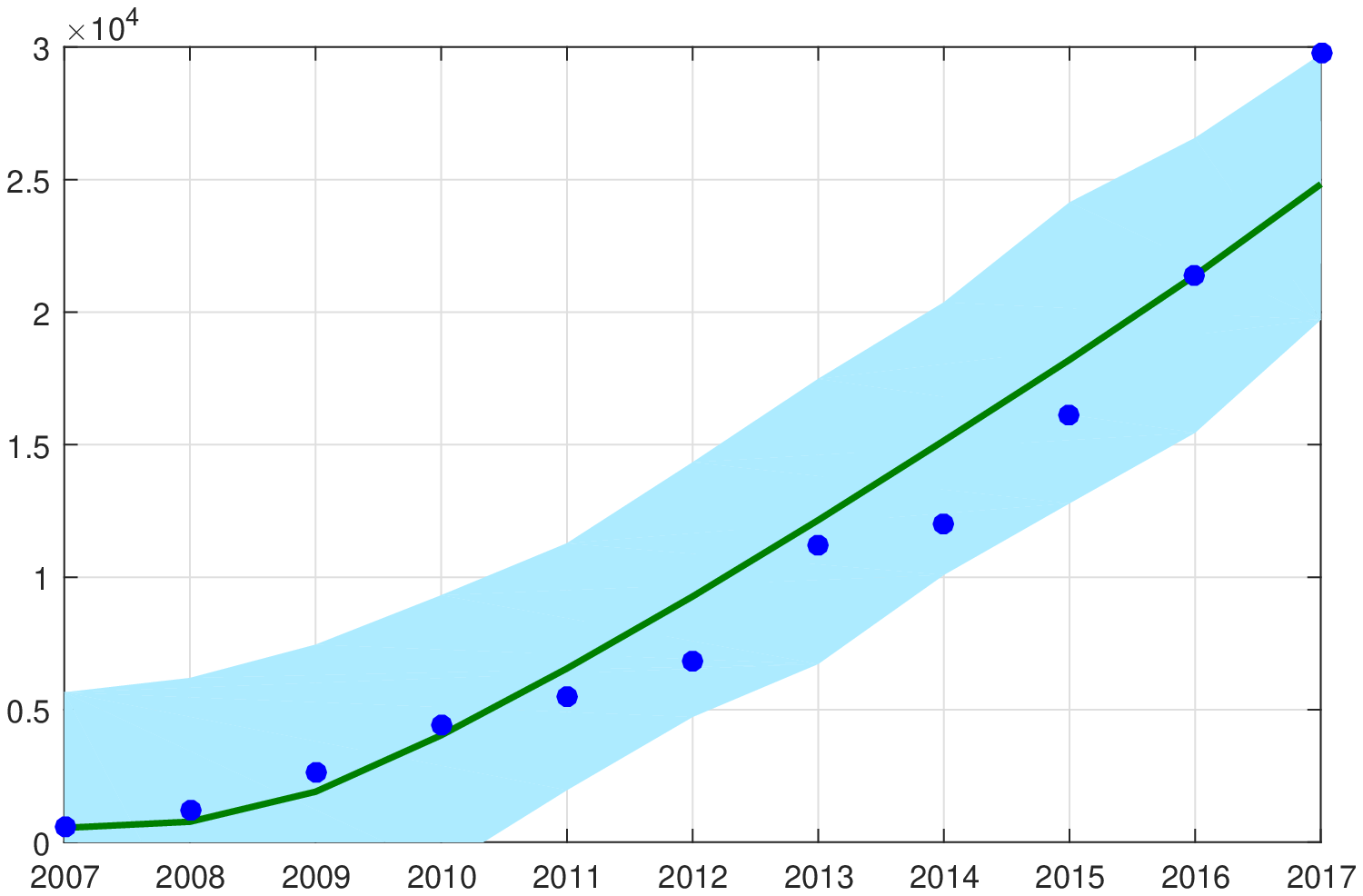}{b}
\includegraphics[width=0.5\textwidth]{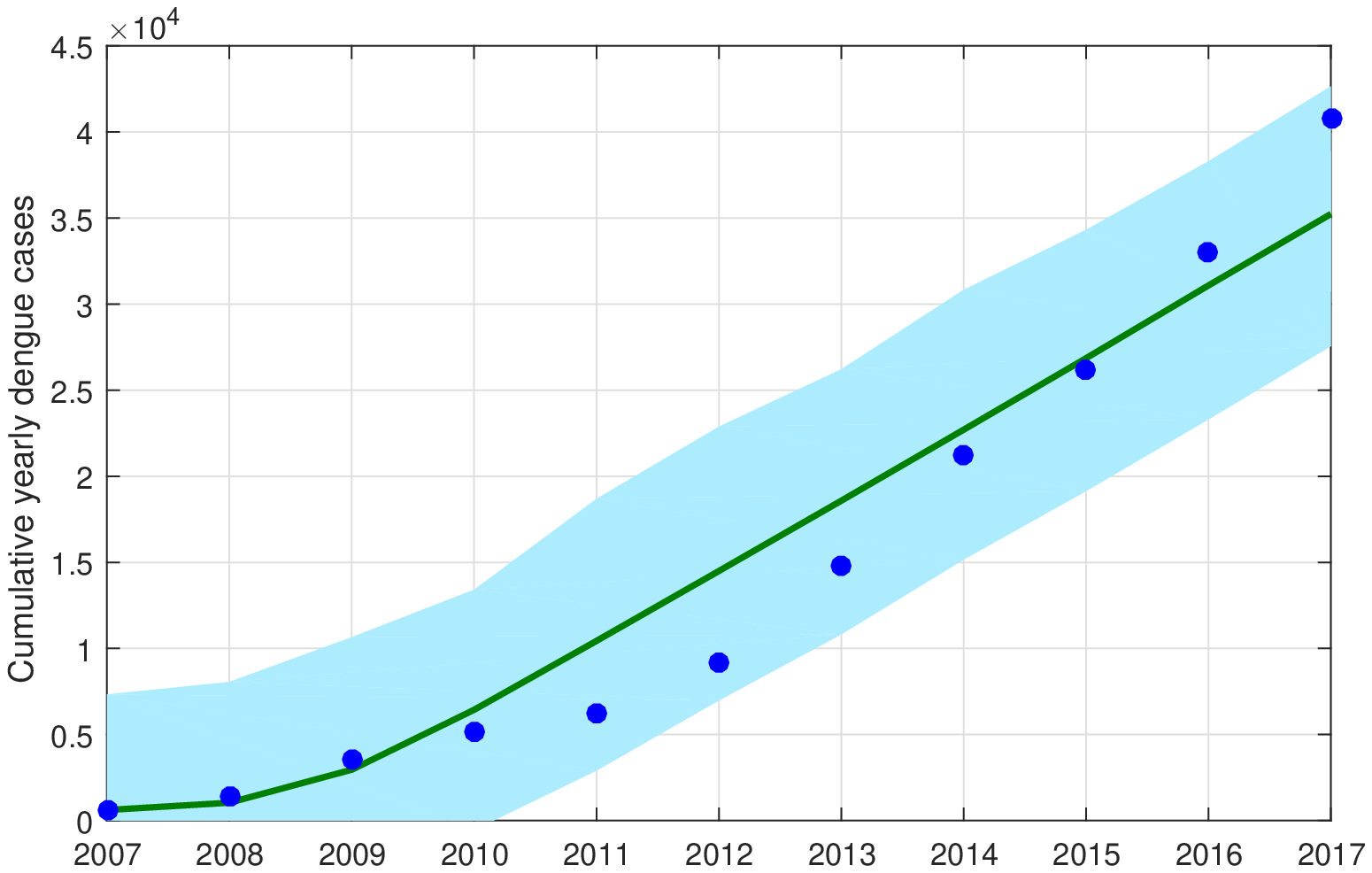}{c}
\includegraphics[width=0.5\textwidth]{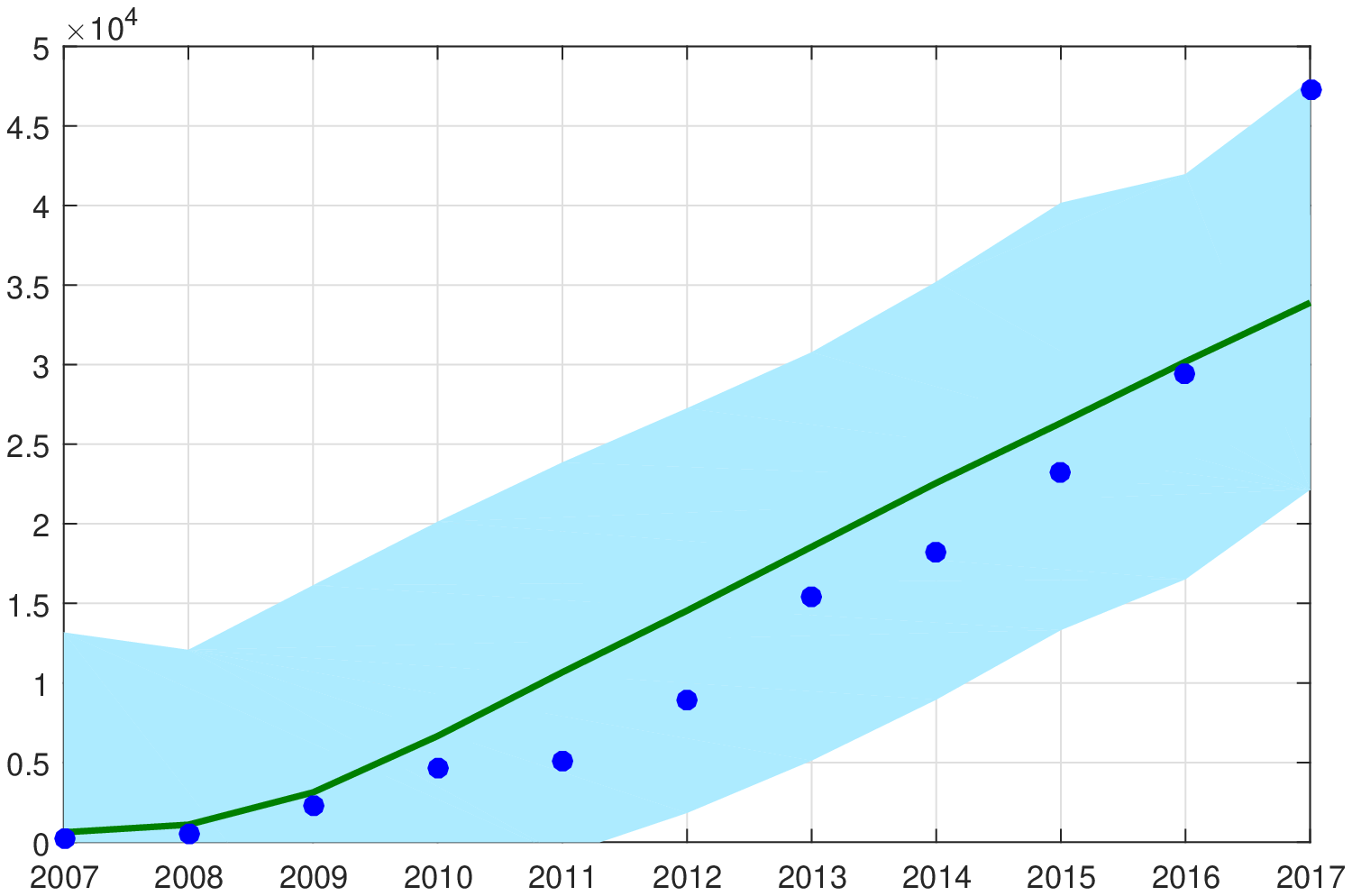}{d}
\caption{Plots of the output of the fitted model (\ref{eq1}) and the
observed cumulative dengue data for (a) Andhra Pradesh, (b)
Rajasthan, (c) Maharashtra and (d) Karnataka. Cumulative cases
(filled blue circle) from the data, and model simulated data (thick
green curve) are plotted with the parameter estimates using
parameter values of Table \ref{table1}.} \label{fig_fitting2}
\end{figure}

We draw initial samples of $\widehat{\theta}$ using the LHS
technique, then obtain an estimate of each sample by finding local
minima of $SS(\widehat{\theta})$ using Nonlinear Least-Square
techniques. The lowest value of $SS(\widehat{\theta})$ is found and
the corresponding $\widehat{\theta}$ is chosen as the initial guess
in the MCMC Toolbox \cite{marko2006}. Further, the convergence of
chain is also confirmed using the Geweke’s Z-scores, Table
\ref{table2}. From the table, we see that the biting rates of
mosquitoes, $\beta$, are higher in Andhra Pradesh, Rajasthan,
Gujarat and Maharashtra. High rates of biting rate in these states
may be due to large number of mosquitoes populations and/or lack of
personal protection. Moreover, the rate of hospitalization and/or notification of symptomatic humans, $\eta$, are estimated to be lower for these four states. These observations indicate that people in these four states may have less awareness about the disease.

\begin{center}
\begin{table}[!htbp]
\caption{Estimated parameters of the system (\ref{eq1}) and their
mean values given in 95\% CI for different states of India}
\begin{tabular}{|c|c|c|c|c|c|}
\hline
States & Parameter & Mean value & 95\% Confidence interval & Geweke's Z-score \\
\hline
 Kerala & $\beta$ & 53.347 & 24.516 -- 85.620 & 0.9029 \\
        & $\eta$ & 0.0195 & 0.0018 -- 0.1459 & 0.8079 \\ \hline
 Delhi & $\beta$ & 42.885 & 26.2043 -- 65.8082 & 0.8681 \\
       & $\eta$ & 0.0143 & 0.0025 -- 0.0579 & 0.8782 \\ \hline
 Gujarat & $\beta$ & 62.224 & 46.8378 -- 80.5926 & 0.9685 \\
         & $\eta$ & 0.0042 & 0.0013 -- 0.0112 & 0.8502 \\ \hline
 West Bengal & $\beta$ & 47.177 & 17.4763 -- 82.1461 & 0.8739 \\
             & $\eta$ & 0.029 & 0.0014 -- 0.1780 & 0.6305 \\ \hline
 Andhra Pradesh & $\beta$ & 67.393 & 50.8408 -- 85.4559 & 0.9769 \\
                & $\eta$ & 0.0025 & 0.0009 -- 0.0065 & 0.7813 \\
                \hline
 Rajasthan & $\beta$ & 62.929 & 39.723 -- 87.823 & 0.8994 \\
            & $\eta$ & 0.0043 & 0.0008 -- 0.0176 & 0.741 \\ \hline
 Maharashtra & $\beta$ & 60.531 & 33.4334 -- 83.6098 & 0.9257 \\
             & $\eta$ & 0.0054 & 0.0011 -- 0.0245 & 0.7928 \\ \hline
 Karnataka & $\beta$ & 43.572 & 19.0592 -- 74.3453 & 0.9694 \\
           & $\eta$ & 0.0126 & 0.0008 -- 0.0645 & 0.8002 \\
 \hline
\end{tabular}
\label{table2}
\end{table}
\end{center}

\section{Impact of ACF on dengue control}\label{control}
\begin{figure}[h]
\includegraphics[width=1.0\textwidth]{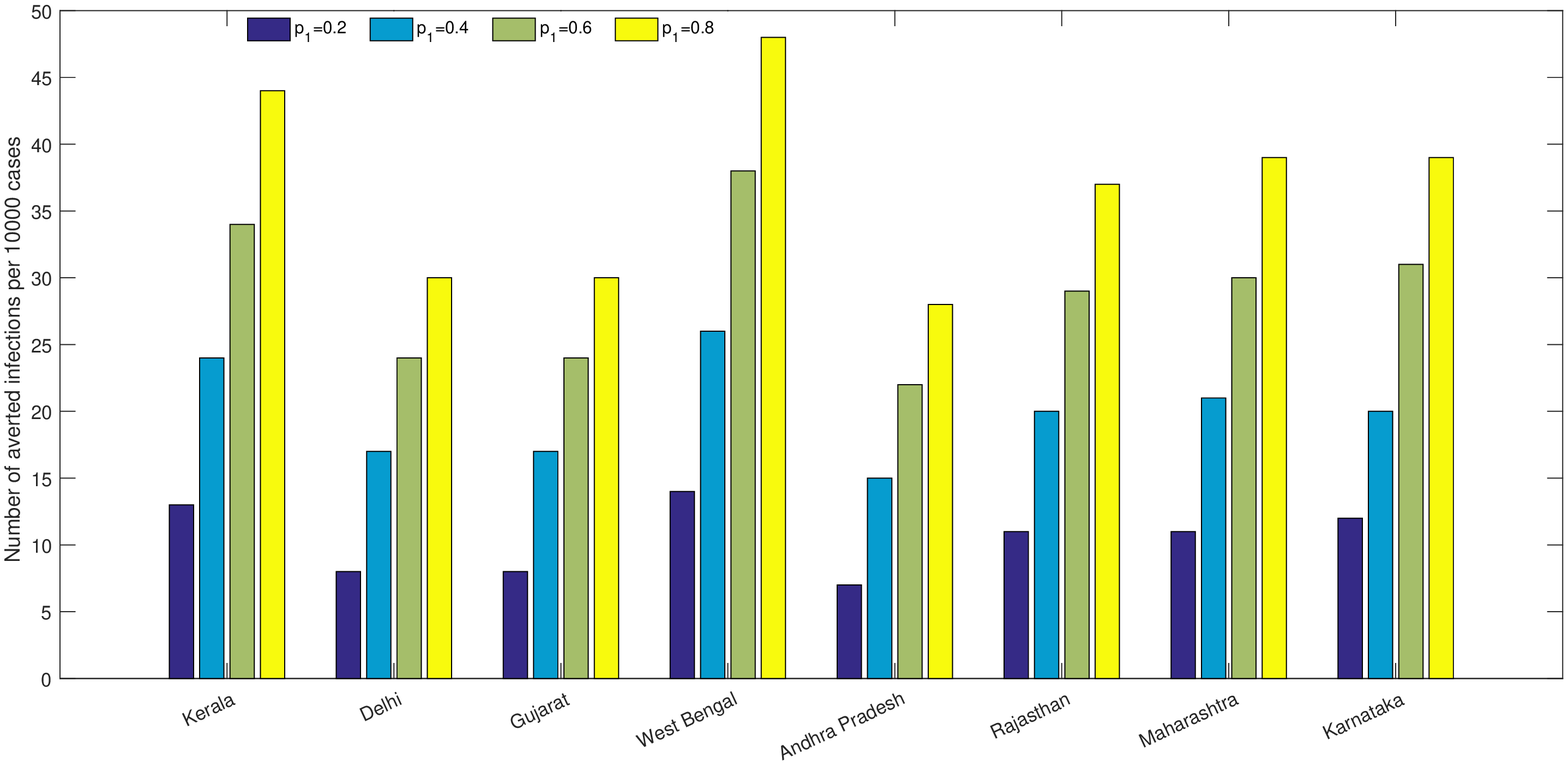}{a}\\
\includegraphics[width=1.0\textwidth]{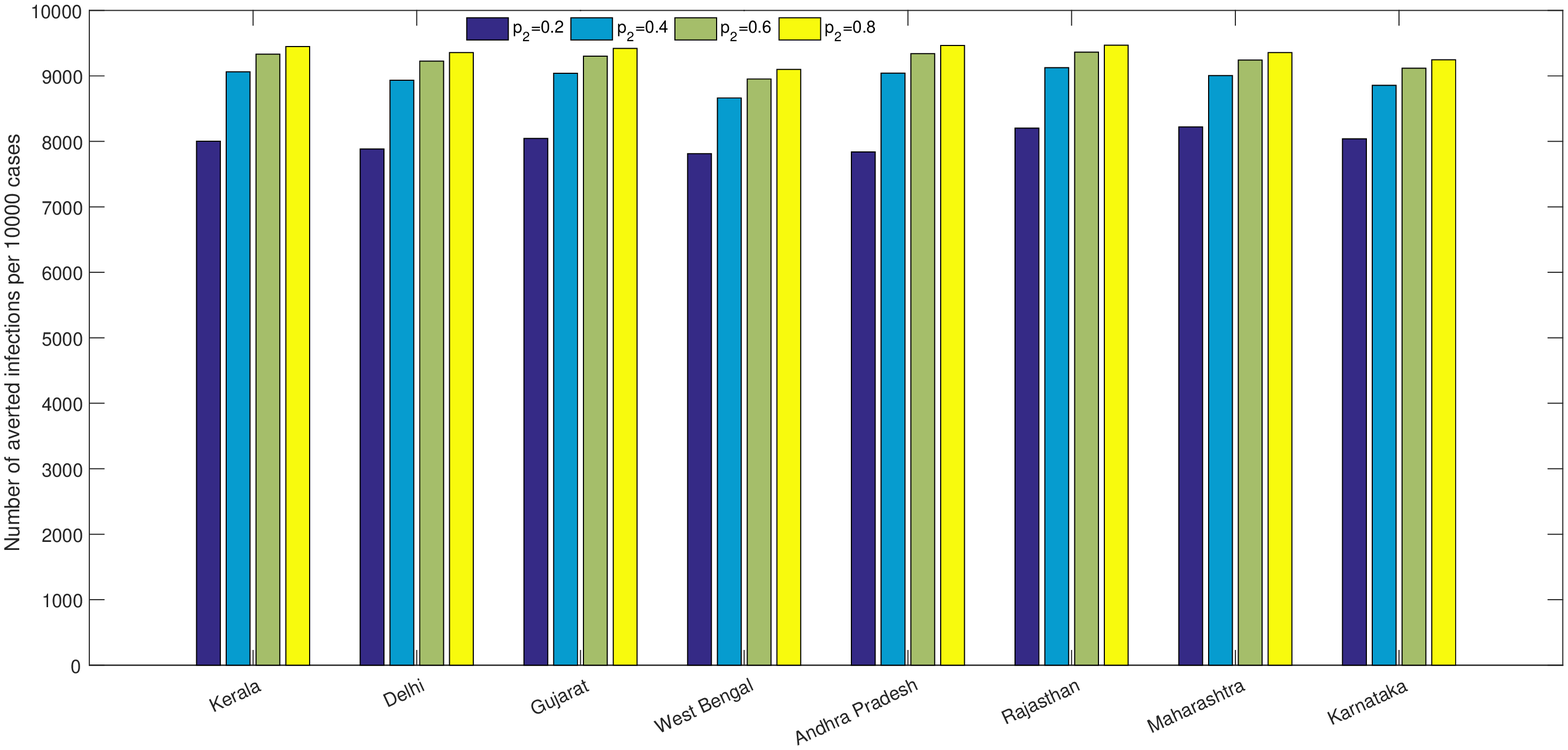}{b}
\caption{Effect of ACF on the total number of new infected humans.}
\label{effect_ACF}
\end{figure}

Active case finding for dengue patients is the systematic
identification of people with suspected dengue, in a predetermined
target area by using tests (such as SD Bioline Dengue Duo Rapid Test
Kit) at a regular basis. The positive ones should be
hospitalized immediately for treatment or the person should be kept
in a mosquito-free environment to avoid secondary infection.
However, the results from sensitivity analysis suggest that $p_2$ is
more effective than $p_1$ in terms of case reduction. Now we
quantify the impacts of these two parameters on the percentage
reduction of dengue cases in the eight states. Using the estimated
parameters (see Table \ref{table2}) for each state we predict total
dengue cases in the years 2018--2050. The base cases were determined
by simulating the model without ACF parameters. For different values
of $p_1$ and $p_2$, the case reduction in total dengue cases is
depicted in Fig. \ref{effect_ACF}.

All of the eight states of India shows similar patterns in case
reduction by the ACF intervention. Note that the scales of averted
cases is different in Fig. \ref{effect_ACF}(a) and Fig.
\ref{effect_ACF}(b). The reason behind this is that the asymptomatic
individuals are less infectious as compared to symptomatic humans.
Maximum employment of ACF of asymptomatic individuals will cause
$0.44\%$, $0.30\%$, $0.30\%$, $0.48\%$, $0.28\%$, $0.37\%$, $0.39\%$
and $0.39\%$ increase in total averted averted cases in Kerala,
Delhi, Gujarat, West Bengal, Andhra Pradesh, Rajasthan, Maharashtra
and Karnataka, respectively. On the other hand maximum employment of
ACF of symptomatic individuals will cause $94.49\%$, $93.57\%$,
$94.19\%$, $90.99\%$, $94.64\%$, $94.69\%$, $93.57\%$ and $92.46\%$
in Kerala, Delhi, Gujarat, West Bengal, Andhra Pradesh, Rajasthan,
Maharashtra and Karnataka, respectively. These results indicate that
$p_2$ is more effective in reducing averted cases and the policy
makers should focus on this strategy to achieve maximum number of
averted dengue cases.

\section{Conclusion and discussion}\label{conclusion}
In this article, we formulated a compartmental ODE model for dengue.
The model included class of hospitalized and/or notified humans who cannot transmit dengue as they
are kept in a mosquito free environment. We showed positivity and boundedness of the
solutions of the system (\ref{eq1}). System
(\ref{eq1}) has a unique disease-free equilibrium which is locally
asymptotically stable if $\mathcal{R}_0<1$ and unstable if
$\mathcal{R}_0>1$. We found that the parameters $\eta$, $\mu_v$,
$p_1$ and $p_2$ have negative effects on $\mathcal{R}_0$ while the
parameter $\beta$ have positive effect (see Section
\ref{sensitivity1}). From the normalized forward sensitive indices
of $\mathcal{R}_0$ and the semi-sensitivity solutions of total
infective populations, we observed that $p_2$ is more effective than
$p_1$ in reducing the disease burden. Therefore, the health care
organizations should pay more attention towards the active case
finding of symptomatic individuals in comparison to that of
asymptomatic.

System (\ref{eq1}) is calibrated using yearly data of dengue from
eight different states of India for the years 2007--2017. Model
fitting with yearly new dengue cases is depicted in Figs.
\ref{fig_fitting1} and \ref{fig_fitting2}, and the 95\% confidence
intervals of the estimated parameters are given in Table
\ref{table2}. The estimated values of biting rate of mosquitoes and rate of hospitalization and/or notification show that Andhra Pradesh, Rajasthan,
Gujarat and Maharashtra are at higher risk of future outbreaks. Moreover, using these parameters, we computed the number of
cases averted by employing ACF in eight different states of India.
It is observed that all of the eight states show similar trend of
case reduction by ACF. From Fig. \ref{effect_ACF}, one can easily
note that ACF of symptomatic individuals will have significant
effect on dengue case reduction. On the other hand, ACF of
asymptomatic humans will avert comparatively less number of dengue
cases. However, it is well established that most of the dengue cases
are asymptomatic. Therefore, asymptomatic humans play an important
role in the persistence of dengue in the community. Complete
eradication of the disease will be difficult unless we control the
asymptomatic individuals. This indicates that ACF of asymptomatic
individuals is not negligible in the long run.

Currently, ACF has been used as an effective control strategy
against tuberculosis (TB) in India \cite{Mandal2015,Nagaraja2017}.
Around 20 million people were tested and a large number of persons were detected positive in the year 2013-2014 \cite{Prasad2016}.
The Revised National TB Control Programme (RNTCP) has decided to
implement ACF for TB in 552 districts of India among high prevalence
area from 2017 onwards as part of its latest national strategic plan
\cite{Nagaraja2017}. Recent studies showed that implementation of
ACF in India can eliminate TB \cite{Nagaraja2017,Prasad2016}.
Results of our study shows similar effects of ACF on dengue control
in India. The healthcare agencies should focus on the areas with
most dengue cases and employ ACF in order to reduce the raising
number of dengue cases in the country. In addition, quite a huge number of missing cases reside in endemic areas of India \cite{Das2017,Kakkar2012}.
ACF will definitely help to fill up the gap of missing cases. Healthcare agencies should identify high-risk target areas and provide proper resources to run the ACF programme smoothly. Due to the twofold benefits, we recommend that along with existing control measures (personal
protection and mosquitoes control), the healthcare organizations
must focus on ACF, which plays a plausible role in reducing the
number of dengue cases to a low endemic equilibrium level in the
endemic states of India.

In future research, we may add host heterogeneity to our
transmission model in order to better understand the impact of the
diffusion of humans \cite{Stolerman2015}. The parameters such as
biting rate depends on climatic factors \cite{Bartley2002},
therefore adding seasonal effects to our model will make it more
realistic. Moreover, ACF strategy can be compared with recent
mosquito control strategies (Ovitraps \cite{Paz-Soldan2016}, virus
supressing Wolbachia infection \cite{Turelli2017} and sterile insect
technique \cite{Mishra2018}) to understand the potential of these
control interventions on the disease eradication.

\section*{Acknowledgement}
The research work of Indrajit Ghosh is supported by University Grants Commission, Government of India, New Delhi in the form of senior research fellowship.
Pankaj Kumar Tiwari is thankful to University Grants Commissions, New Delhi,
India for providing financial support in form of D. S. Kothari
post-doctoral fellowship (No.F.4-2/2006 (BSR)/MA/17-18/0021).

\section*{Appendix A}

System (\ref{eq1}) can be rewritten as
\begin{eqnarray*}
\frac{dX}{dt}=CX+D,
\end{eqnarray*}
$X=[S_v,E_v,I_v,S_{h1},S_{h2},E_h,A_h,I_h,P_h,R_h]^T$ and
$$C=\left[\begin{array}{cccccccccc}
d_1 & 0 & 0 & 0 & 0 & 0 & 0 & 0 & 0 & 0\\
d_2 & d_3 & 0 & 0 & 0 & 0 & 0 & 0 & 0 & 0\\
0 & \gamma_v & -\mu_v & 0 & 0 & 0 & 0 & 0 & 0 & 0\\
0 & 0 & 0 & d_4 & 0 & 0 & 0 & 0 & 0 & 0\\
0 & 0 & 0 & 0 & d_5 & 0 & 0 & 0 & 0 & 0\\
0 & 0 & d_6 & 0 & 0 & d_7 & 0 & 0 & 0 & 0\\
0 & 0 & 0 & 0 & 0 & \rho\gamma_h & d_8 & 0 & 0 & 0\\
0 & 0 & 0 & 0 & 0 & (1-\rho)\gamma_h & 0 & d_9 & 0 & 0\\
0 & 0 & 0 & 0 & 0 & 0 & p_1 & d_{10} & d_{11} & 0\\
0 & 0 & 0 & 0 & 0 & 0 & q_1 & q_2 & q_3 & -\mu_h
\end{array}
\right],$$
where
\begin{eqnarray*}
&&d_1=-\left[\mu_v+\beta\alpha_h\left(\frac{I_h+\lambda
A_h}{N_h}\right)\right], \ d_2=\beta\alpha_h\left(\frac{I_h+\lambda
A_h}{N_h}\right), \ d_3=-(\gamma_v+\mu_v),\\
&&d_4=-\left[\mu_h+\beta\alpha_v\left(\frac{I_v}{N_h}\right)\right],
\
d_5=-\left[\mu_h+\beta\theta\alpha_vI_v\left(\frac{S_{h2}}{N_h}\right)\right],
\ d_6=\beta\alpha_v\left(\frac{S_{h1}+\theta S_{h2}}{N_h}\right),\\
&&d_7=-(\gamma_h+\mu_h), \ d_8=-(\mu_h+q_1+p_1), \
d_9=-(\mu_h+\eta+q_2+p_2), \ d_{10}=\eta+p_2, \
d_{11}=-(q_3+\mu_h+\delta).
\end{eqnarray*}

The vector $D=\left[\pi_v,0,0,r\pi_h,(1-r)\pi_h,0,0,0,0,0\right]^T$
is positive. Note that all the off-diagonal entries of $C(X)$ are
nonnegative. Therefoe, the matrix $C(X)$ is Metzler for all $X\in
\mathbb{R}^{10}_+$. Thus, system (\ref{eq1}) is positively invariant
in $\mathbb{R}^{10}_+$ \cite{Abate2009}. Hence, all trajectories of
the system (\ref{eq1}) which originate from an initial state in
$\mathbb{R}^{10}_+$ confine therein forever.

Summing up the first three equations of system (\ref{eq1}), we get
\begin{eqnarray*}
\frac{dN_v}{dt}=\pi_v-\mu_vN_v.
\end{eqnarray*}

Using a standard comparison theorem \cite{Lakshmikantham1989}, we have $\displaystyle 0 \leq N_v(t) \leq
\frac{\pi_v}{\mu_v}+\left(N_v(0)-\frac{\pi_v}{\mu_v}\right)e^{-\frac{\pi_v}{\mu_v}}.$\
Thus, as $t \rightarrow \infty$, $\displaystyle 0 \leq N_v(t)
\leq \frac{\pi_v}{\mu_v},$\ we have for any $t>0$, $0 \leq
N_v(t) \leq Z_1$, where $\displaystyle
Z_1=\max\left\{\frac{\pi_v}{\mu_v},N_v(0)\right\}.$\

Assume that $\displaystyle
Z_2=\max\left\{\frac{r\pi_h}{\mu_h},S_{h1}(0)\right\}.$\ Then $0
\leq S_{h1} \leq Z_2$. Similarly, let $\displaystyle
Z_3=\max\left\{\frac{(1-r)\pi_h}{\mu_h},S_{h2}(0)\right\}.$\ Then $0
\leq S_{h2} \leq Z_3$.

By adding the last seven equations of the system (\ref{eq1}),
we get
\begin{eqnarray*}
\frac{dN_h}{dt}=\pi_h-\mu_hN_h-\delta P_h \leq \pi_h-\mu_hN_h.
\end{eqnarray*}

Assume that $\displaystyle
Z_4=\max\left\{\frac{\pi_h}{\mu_h},N_h(0)\right\}.$\ Then $N_h \leq
Z_4.$\ Also, $\displaystyle \frac{dN_h}{dt} \geq
\pi_h-(\mu_h+\delta)N_h.$\ Assume that $\displaystyle
Z_5=\min\left\{\frac{\pi_h}{\mu_h+\delta},N_h(0)\right\}.$\ Then
$N_h \geq Z_5$. Thus, we have $\displaystyle Z_5 \leq N_h \leq
Z_4.$\

Therefore, all feasible solutions of the system (\ref{eq1}) enter
the region $\Omega$ implying that the region is an attracting set.

\section*{Appendix B}

Jacobian of system \eqref{eq1} at the equilibrium $E_0$ is {\tiny
\[J_{E_0}=
   \left[
    \begin{array}{cccccccccc}
           -\mu_v & 0 & 0 & 0 & 0 & 0 & \displaystyle -\frac{\beta\alpha_h\lambda\pi_v\mu_h}{\pi_h\mu_v} & \displaystyle -\frac{\beta\alpha_h\pi_v\mu_h}{\pi_h\mu_v} & 0 & 0\\
           0 & -(\gamma_v+\mu_v) & 0 & 0 & 0 & 0 & \displaystyle \frac{\beta\alpha_h\lambda\pi_v\mu_h}{\pi_h\mu_v} & \displaystyle \frac{\beta\alpha_h\pi_v\mu_h}{\pi_h\mu_v} & 0 & 0\\
           0 & \gamma_v & -\mu_v & 0 & 0 & 0 & 0 & 0 & 0 & 0\\
           0 & 0 & -\beta\alpha_v r & -\mu_h & 0 & 0 & 0 & 0 & 0 & 0\\
           0 & 0 & -\beta\theta\alpha_v(1-r) & 0 & -\mu_h & 0 & 0 & 0 & 0 & 0\\
           0 & 0 & \beta\alpha_v\{r+\theta(1-r)\} & 0 & 0 & -(\gamma_h+\mu_h) & 0 & 0 & 0 & 0\\
           0 & 0 & 0 & 0 & 0 & \rho\gamma_h & -(\mu_h+q_1+p_1) & 0 & 0 & 0\\
           0 & 0 & 0 & 0 & 0 & (1-\rho)\gamma_h & 0 & -(\mu_h+\eta+q_2+p_2) & 0 & 0\\
           0 & 0 & 0 & 0 & 0 & 0 & p_1 & \eta+p_2 & -(\mu_h+\delta+q_3) & 0\\
           0 & 0 & 0 & 0 & 0 & 0 & q_1 & q_2 & q_3 & -\mu_h
    \end{array}
    \right].\]}

Five eigenvalues of the matrix $J_{E_0}$ are $-\mu_v$, $-\mu_h$ (of multiplicity 3) and
$-(\mu_h+\delta+q_3)$, and other five are given by the roots of the equation
\begin{eqnarray}\label{am20}
&&\rho^5+A_1\rho^4+A_2\rho^3+A_3\rho^2+A_4\rho+A_5=0,
\end{eqnarray}
where
\begin{eqnarray*}
&&A_1=2\mu_v+3\mu_h+\eta+\gamma_v+\gamma_h+q_1+q_2+p_1+p_2,\\
&&A_2=(\mu_h+\eta+q_2+p_2)(2\mu_v+2\mu_h+\gamma_v+\gamma_h+q_1+p_1)+(\mu_h+q_1+p_1)(2\mu_v+\mu_h+\gamma_v+\gamma_h)+\mu_v(\gamma_v+\mu_v)\\
&&\hskip 1cm +(\gamma_h+\mu_h)(\gamma_v+2\mu_v),\\
&&A_3=(\mu_h+\eta+q_2+p_2)\{(\mu_h+q_1+p_1)(2\mu_v+\mu_h+\gamma_v+\gamma_h)+\mu_v(\gamma_v+\mu_v)+(\gamma_h+\mu_h)(\gamma_v+2\mu_v)\}\\
&&\hskip 1cm +\mu_v(\gamma_v+\mu_v)(\gamma_h+\mu_h)
+(\mu_h+q_1+p_1)\{\mu_v(\gamma_v+\mu_v)+(\gamma_h+\mu_h)(\gamma_v+2\mu_v)\},\\
&&A_4=(\mu_h+\eta+q_2+p_2)(\mu_h+q_1+p_1)\{\mu_v(\gamma_v+\mu_v)+(\gamma_h+\mu_h)(\gamma_v+2\mu_v)\}\\
&&\hskip 1cm +\frac{\beta^2\alpha_v\alpha_h\gamma_v\gamma_h\pi_v\mu_h\{r+\theta(1-r)\}}{\mu_v\pi_h}\left[(\mu_h+\eta+q_2+p_2)\frac{\lambda\rho}{\mu_h+q_1+p_1}
+(\mu_h+q_1+p_1)\frac{(1-\rho)}{\mu_h+\eta+q_2+p_2}\right]\\
&&\hskip 1cm +\mu_v(\mu_h+\gamma_h)(\mu_v+\gamma_v)(2\mu_h+\eta+q_1+q_2+p_1+p_2)(1-R^2_0),\\
&&A_5=\mu_v(\gamma_v+\mu_v)(\gamma_h+\mu_h)(\mu_h+q_1+p_1)(\mu_h+\eta+q_2+p_2)(1-R^2_0).
\end{eqnarray*}

Clearly, all roots of equation (\ref{am20}) are either negative or have
negative real parts if $R_0<1$. Thus, the disease-free equilibrium
$E_0$ is locally asymptotically stable if $R_0<1$ and unstable if
$R_0>1$.

\section*{Appendix C}

Consider the following positive definite function
\begin{eqnarray}\label{eqg2}
&&V=\frac{1}{2}[(S_v-S^*_v)^2+(E_v-E^*_v)^2+(I_v-I^*_v)^2+(N_h-N^*_h)^2+(S_{h2}-S^*_{h2})^2\nonumber\\
&&\hskip 1cm +(E_h-E^*_h)^2+(A_h-A^*_h)^2+(I_h-I^*_h)^2+(P_h-P^*_h)^2+(R_h-R^*_h)^2].
\end{eqnarray}

Differentiating equation (\ref{eqg2}) with respect to time `$t$' along
the solution trajectories of system ({\ref{eqg1}) and rearranging the terms, we get
\begin{eqnarray*}
&&\frac{dV}{dt}=-\left[\mu_v+\beta\alpha_h\left(\frac{I^*_h+\lambda A^*_h}{N^*_h}\right)\right](S_v-S^*_v)^2-[\gamma_v+\mu_v](E_v-E^*_v)^2-[\mu_v](I_v-I^*_v)^2\\
&&\hskip 1cm -[\mu_h](N_h-N^*_h)^2-\left[\frac{\theta\beta\alpha_vI^*_v}{N^*_h}+\mu_h\right](S_{h2}-S^*_{h2})^2
-\left[\frac{\beta\alpha_vI_v}{N_h}+\gamma_h+\mu_h\right](E_h-E^*_h)^2\\
&&\hskip 1cm -[\mu_h+q_1+p_1](A_h-A^*_h)^2-[\mu_h+\eta+q_2+p_2](I_h-I^*_h)^2-[q_3+\mu_h+\delta](P_h-P^*_h)^2-[\mu_h](R_h-R^*_h)^2\\
&&\hskip 1cm -\frac{\beta\alpha_hS_v(I_h+\lambda A_h)}{N_hN^*_h}(S_v-S^*_v)(N_h-N^*_h)-\frac{\beta\alpha_hS_v}{N^*_h}(S_v-S^*_v)(I_h-I^*_h)
-\frac{\beta\lambda\alpha_h}{N^*}(S_v-S^*_v)(A_h-A^*_h)\\
&&\hskip 1cm +\frac{\beta\alpha_h(I^*_h+\lambda A^*_h)}{N^*_h}(S_v-S^*_v)(E_v-E^*_v)+\frac{\beta\alpha_hS_v(I_h+\lambda A_h)}{N_hN^*_h}(E_v-E^*_v)(N_h-N^*_h)\\
&&\hskip 1cm +\frac{\beta\alpha_hS_v}{N^*_h}(E_v-E^*_v)(I_h-I^*_h)+\frac{\beta\lambda\alpha_hS_v}{N^*}(E_v-E^*_v)(A_h-A^*_h)+[\gamma_v](E_v-E^*_v)(I_v-I^*_v)\\
&&\hskip 1cm -[\delta](N_h-N^*_h)(P_h-P^*_h)-\frac{\beta\theta\alpha_vI_vS_{h2}}{N_hN^*_h}(N_h-N^*_h)(S_{h2}-S^*_{h2})
-\frac{\theta\beta\alpha_vS_{h2}}{N^*_h}(I_v-I^*_v)(S_{h2}-S^*_{h2})\\
&&\hskip 1cm +\frac{\beta\alpha_v}{N_hN^*_h}\{I^*_v(S^*_{h2}+E^*_h+A^*_h+I^*_h+P^*_h+R^*_h)+\theta S_{h2}I_v\}(N_h-N^*_h)(E_h-E^*_h)\\
&&\hskip 1cm +\beta\alpha_v\left[\frac{N_h-S^*_{h2}-E^*_h-A^*_h-I^*_h-P^*_h-R^*_h}{N_h}+\frac{\theta S_{h2}}{N^*_h}\right](I_v-I^*_v)(E_h-E^*_h)\\
&&\hskip 1cm +\beta\alpha_v\left[\frac{\theta I^*_v}{N^*_h}-\frac{I_v}{N_h}\right](S_{h2}-S^*_{h2})(E_h-E^*_h)-\frac{\beta\alpha_vI_v}{N_h}(E_h-E^*_h)(P_h-P^*_h)\\
&&\hskip 1cm -\frac{\beta\alpha_vI_v}{N_h}(E_h-E^*_h)(R_h-R^*_h)+\left[\rho\gamma_h-\frac{\beta\alpha_vI_v}{N_h}\right](E_h-E^*_h)(A_h-A^*_h)\\
&&\hskip 1cm +\left[(1-\rho)\gamma_h-\frac{\beta\alpha_vI_v}{N_h}\right](E_h-E^*_h)(I_h-I^*_h)+[p_1](A_h-A^*_h)(P_h-P^*_h)+[\eta+p_2](I_h-I^*_h)(P_h-P^*_h)\\
&&\hskip 1cm
+[q_1](A_h-A^*_h)(R_h-R^*_h)+[q_2](I_h-I^*_h)(R_h-R^*_h)+[q_3](P_h-P^*_h)(R_h-R^*_h).
\end{eqnarray*}

Inside the region of attraction $\Omega$, $\displaystyle \frac{dV}{dt}$\ can be made negative definite provided the inequalities (\ref{eqgl3})$-$(\ref{eqgl8}) hold.

\end{document}